\documentclass[fleqn,usenatbib]{aa}
\usepackage{graphicx}
\usepackage{general}
\usepackage{amssymb,amsmath}
\usepackage[varg]{txfonts}
\usepackage[pdftex,colorlinks=true,citecolor=blue,linkcolor=orange]{hyperref}

\newcommand{\hba}{Paper~I}
\newcommand{\nqnhp}{\ce{N$^{15}$NH+}}
\newcommand{\qnnhp}{\ce{$^{15}$NNH+}}
\newcommand{\rr}{\foun/\fifn}
\newcommand{\rrr}{\ensuremath{{\cal R}}}
\newcommand{\rtot}{\ensuremath{{\cal R}_{\rm tot}}}
\newcommand{\rx}[1]{\ensuremath{{\cal R(\ce{#1})}}}

\newcommand{\agr}{\ensuremath{{a_{\rm gr}}}}
\newcommand{\acore}{\ensuremath{{a_{\rm c}}}}
\renewcommand{\rho}{\varrho}
\newcommand{\rhocore}{\ensuremath{{\rho_{\rm core}}}}
\newcommand{\rhoenv}{\ensuremath{{\rho_{\rm env}}}}
\newcommand{\rhogr}{\ensuremath{{\rho_{\rm c}}}}
\newcommand{\taufo}{\ensuremath{\tau_{\rm fo}}}
\newcommand{\tauff}{\ensuremath{\tau_{\rm ff}}}
\newcommand{\taufr}{\ensuremath{\tau_{\rm fr}}}
\newcommand{\vther}{\ensuremath{v_{\rm th}}}
\newcommand{\mamu}{\ensuremath{\tilde{m}}}

\newcommand{\alico}{\textsc{Alico}}

\defcitealias{terzieva2000}{TH00}
\defcitealias{roueff2015}{R15}

\begin{document}

\title{Depletion and fractionation of nitrogen in collapsing cores}
\author{%
  P. Hily-Blant\inst{1}\and
  G. Pineau des For\^ets\inst{2,3}\and
  A. Faure\inst{1}\and
  D. R. Flower\inst{4}
}

\institute{%
	Univ. Grenoble Alpes, CNRS, IPAG, 38000 Grenoble, France
	\email{pierre.hily-blant@univ-grenoble-alpes.fr},
	\and
	Université Paris-Saclay, CNRS, Institut d’Astrophysique Spatiale, 91405, Orsay, France
	\and
	Observatoire de Paris, PSL university, Sorbonne Université, CNRS, LERMA, 75014, Paris, France
	\and
	Physics Department, The University, Durham DH1 3LE, UK
}

\date{}

\abstract{
	Measurements of the nitrogen isotopic ratio in Solar System comets show a constant value, $\approx$140, which is three times lower than the protosolar ratio, a highly significant difference that remains unexplained. Observations of static starless cores at early stages of collapse confirm the theoretical expectation that nitrogen fractionation in interstellar conditions is marginal for most species. Yet, observed isotopic ratios in N$_2$H$^+$ are at variance with model predictions. These gaps in our understanding of how the isotopic reservoirs of nitrogen evolve, from interstellar clouds to comets, and, more generally, to protosolar nebulae, may have their origin in missing processes or misconceptions in the chemistry of interstellar nitrogen. So far, theoretical studies of nitrogen fractionation in starless cores have addressed the quasi-static phase of their evolution such that the effect of dynamical collapse on the isotopic ratio is not known. In this paper, we investigate the fractionation of $^{14}$N and $^{15}$N during the gravitational collapse of a pre-stellar core through gas-phase and grain adsorption and desorption reactions. The initial chemical conditions, which are obtained in steady state after typically a few Myr, show low degrees of fractionation in the gas phase, in agreement with earlier studies. However, during collapse, the differential rate of adsorption of $^{14}$N- and $^{15}$N-containing species onto grains results in enhanced $^{15}$N:$^{14}$N ratios, in better agreement with the observations. Furthermore, we find differences in the behavior, with increasing density, of the isotopic ratio in different species. We find that the collapse must take place on approximately one free-fall timescale, based on the CO abundance profile in L183. Various chemical effects that bring models into better agreement with observations are considered. Thus, the observed values of $^{14}$N$_2$H$^+$:N$^{15}$NH$^+$ and $^{14}$N$_2$H$^+$:$^{15}$NNH$^+$ could be explained by different temperature dependences of the rates of dissociative recombination of these species. We also study the impact of the isotopic sensitivity of the charge-exchange reaction of N$_2$ with He$^+$ on the fractionation of ammonia and its singly deuterated analog and find significant depletion in the $^{15}$N variants. However, these chemical processes require further experimental and theoretical investigations, especially at low temperature. These new findings, such as the depletion-driven fractionation, may also be relevant to the dense, UV-shielded regions of protoplanetary disks.}


\maketitle

\section{Introduction}
\label{sec:intro}

\begin{figure*}
	\centering
	\includegraphics[width=\hsize]{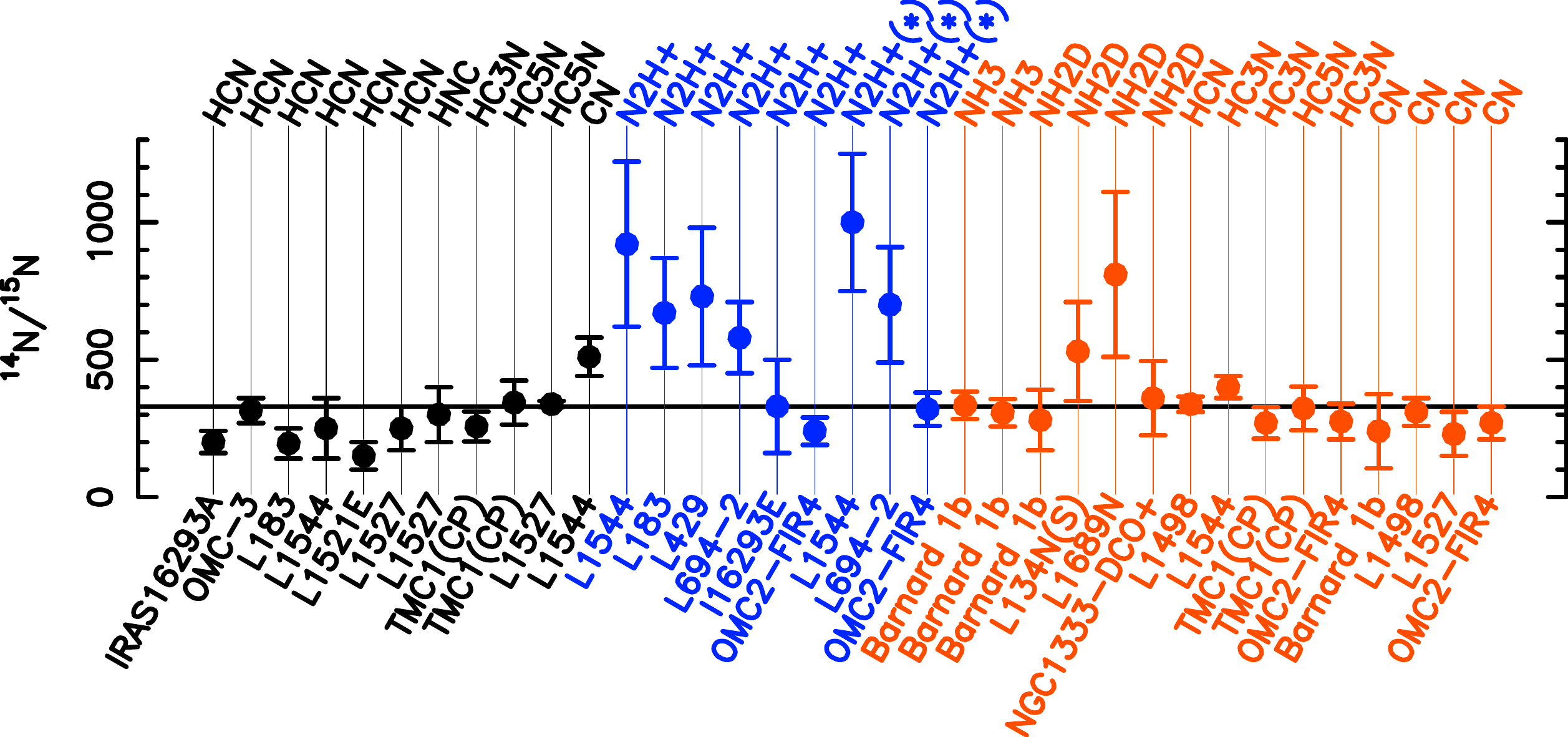}
	\caption{Overview of measurements of the nitrogen isotopic ratio in low- to intermediate-mass cores and Class0. The horizontal line indicates the value of the {bulk} isotopic ratio of nitrogen adopted in this study (\rtot=330). Black dots indicate indirect measurements based on the double-isotope method. Colored points refer to direct measurements, with N$_2$H$^{+}$:N$^{15}$NH$^{+}$ {outlined} in blue. The N$_2$H$^{+}$:$^{15}$NNH$^{+}$ ratios are indicated  by an asterisk. {See \rapp{review} and Table~\ref{tab:review} for details.}}
	\label{fig:review}
\end{figure*}

Protoplanetary disks provide the material out of which planets and primitive cosmic objects, such as meteorites and comets, form, and establishing their chemical composition has become a key objective \citep{vandishoeck2014a, vandishoeck2018}. More specifically, recent years have seen renewed interest in the isotopic ratio of interstellar nitrogen, with a view to establishing the degree to which planetary systems inherit their chemical composition from their parent interstellar clouds \citep{hilyblant2013a}. In this context, the unprecedented sensitivity of the ALMA interferometer plays an important role, in that it enables the nitrogen isotopic ratios to be measured in circumstellar disks. Similarly, the new generation of receivers at the IRAM 30~m telescope has allowed isotopic ratios to be measured in acceptable amounts of time. Isotopic ratios have long been used to trace the history of cosmic objects such as planets \citep[e.g.,][]{marty2017}. In the case of nitrogen, however, isotopic ratios measured in a variety of solar-system objects present a rather confusing picture, with values ranging from 441 in Jupiter and the proto-Sun, down to approximately 50 in micron-sized inclusions in chondrites \citep{bonal2010}. Interestingly, no ratio higher than the protosolar value of 441 has been reported so far in the Solar System. The current state of affairs is that the origin of nitrogen in the Solar System remains elusive \citep{hilyblant2013a, furi2015}. One of the current challenges is to determine the sources of isotopic ratio variations of nitrogen in star-forming regions and protoplanetary disks.

Progress in establishing the origin of nitrogen in the Solar System relies on a combination of theoretical and observational investigations of star-forming regions and protoplanetary disks, with the aim of obtaining an accurate picture of the processes that determine the isotopic ratios of nitrogen-bearing species. In so doing, care must {be} taken when comparing the 4.6~Gyr-old protosolar nebula to present-day star- and planet-forming regions since stellar nucleosynthesis, or the migration of the Solar System from a different galactocentric radius \citep{minchev2013, barbosa2015}, could {change the bulk ratio \rtot}. Indeed, isotopic ratios well below 441, and as low as 270, have been measured in the diffuse \citep{lucas1998} and dense \citep{adande2012, hilyblant2013a, daniel2013} molecular local interstellar medium (ISM). When comparing the ratios in the present-day ISM to the value of 441 for the proto-Sun, it was realized that the above effects---stellar nucleosynthesis or the migration of the Solar System---must be included \citep{hilyblant2017}. Indeed, accurate measurements of nitrogen isotopic ratios in the local ISM performed in recent years {(see \rfig{review} and \rapp{obs})} have been found to be in very good agreement with predictions of the current bulk ratio from galactic chemical evolution models \citep{romano2017}. The isotopic ratio in the local ISM is now proposed to be 330, thus significantly lower than the protosolar value of 441 \citep{hilyblant2017, hilyblant2019}. A low isotopic ratio in the solar neighborhood has received further observational support \citep{kahane2018,colzi2018a}. Yet, the 330 value is still uncertain, and could be as low as $\approx 270$ \citep{lucas1998, adande2012}. This low value of the isotopic ratio is close to the terrestrial value (272), although this is likely a chance agreement rather than indicative of a true chemical link. In this paper, we will adopt \rtot=330 as the {bulk} isotopic ratio.

%In so doing, care must taken when comparing the 4.6~Gyr-old protosolar nebula to present-day star- and planet-forming regions since stellar nucleosynthesis or migration of the Solar System from a different galactocentric radius \citep{minchev2013, barbosa2015} could change the elemental value of the \rr\ ratio. Accurate measurements of nitrogen isotopic ratios in the local interstellar medium (ISM) performed in the recent years (see Table~\ref{tab:review}) have been found in very good agreement with predictions of the current bulk \rr\ ratio from Galactic Chemical Evolution models \citep{romano2017}. The elemental ratio in the local ISM is now proposed to be 330, thus si3gnificantly lower than the protosolar value of 441 \citep{hilyblant2017, hilyblant2019b}. Note however that the 330 value is still uncertain, and could be as low as $\approx 270$ \citep{lucas1998, adande2012}. Nevertheless, a low elemental ratio in the solar neighborhood has received further observational support \citep[e.g.][]{kahane2018,colzi2018a}. We also emphasize that the low value of the elemental ratio is close to the terrestrial of 272, although this is likely a chance agreement rather than indicative of a true chemical link. In this paper, we will adopt 330 as the elemental ratio.

From a theoretical perspective, two processes are known that could lead to fractionation---that is, a deviation of the isotopic ratio in some species, such as HCN/HC\fifn, from the {bulk} value---in star-forming regions and protoplanetary disks. One is isotope-selective photodissociation of N$_2$ \citep{heays2014, heays2017}, which is expected to be effective in environments where UV radiation has a significant impact on chemistry, such as the upper layers of protoplanetary disks. A recent measurement of the $^{14}$N:$^{15}$N ratio in the CN and HCN molecules in the circumstellar disk of the 8~Myr-old T~Tauri object TW~Hya demonstrated that the HCN:HC$^{15}$N ratio increases with distance from the central star \citep{hilyblant2019}. This observational result lends support to the hypothesis that selective photodissociation of N$_{2}$ \citep{heays2014} could be the leading fractionation process in such disks. Disk models predict some enrichment of CN and HCN in $^{15}$N \citep{visser2018} but lead to lower fractionation levels than observed, which could result from several factors, including the UV flux or the dust properties (size distribution, mass fraction). In shielded gas, such as the collapsing clouds considered in the present study, isotope-selective photodissociation is relevant only through the differential dissociation of $^{14}$N$_2$ and $^{15}$N$^{14}$N by the cosmic-ray induced ultraviolet radiation.

The second process is known as chemical mass fractionation \citep{watson1976}: Heavier isotopic forms have lower zero-point energies, $\hbar \omega / 2$, where $\omega \propto \mu ^{-1/2}$ is the characteristic vibrational frequency, and $\mu $ is the reduced mass of the molecule. As a consequence of this energy difference, molecules can become enriched in a heavier isotope, through isotope-exchange reactions,  when the kinetic temperature of the medium is comparable to, or lower than, the energy difference. Whilst this effect is most pronounced in the case of light molecules, such as H$_2$ and its deuterated forms, for which the differences in reduced mass, and hence in zero-point energies, are substantial, the rare isotopes of other elements, such as $^{13}$C and $^{15}$N, can also become enriched when the temperature is sufficiently low \citep{terzieva2000, roueff2015}. It has been proposed that the $^{15}$N-rich nitrogen reservoirs observed in the Solar System could be formed by chemical mass fractionation in the parent interstellar cloud \citep{charnley2002, hilyblant2013a, hilyblant2013b, furuya2018}; but more sophisticated modeling of the observations has called this interpretation of nitrogen fractionation into question \citep{roueff2015}. That chemical fractionation is not efficient, even in cold gas, is also supported by state-of-the-art determinations of the HCN:HC\fifn\ ratio in the L1498 pre-stellar core by \cite{magalhaes2018a}. These authors obtained HCN:HC$^{15}$N=338$\pm$28, which is in harmony with the {bulk} nitrogen isotopic ratio \rtot=330 in the ISM \citep{hilyblant2017} and indicates that chemical mass fractionation is inefficient, at least for HCN.

From the picture drawn above, one might conclude that models and observations agree qualitatively, in that \fifn-rich reservoirs are formed in situ in protoplanetary disks by the action of isotope-selective photodissociation of \ce{N2}, inherited from a non-fractionated interstellar reservoir. However, measurements of the \ce{N2H+} isotopic ratio in dense clouds are a factor of $\approx2-3$ times higher than the {bulk value \rtot=330}, with no clear origin for the discrepancy; this is the most striking observational evidence that our models of nitrogen fractionation are still incomplete. Attempts to reproduce the observed N$_2$H$^{+}$:N$^{15}$NH$^{+}$ column density ratio with chemical models have failed so far \citep{roueff2015, wirstrom2018, loison2019}. It may be noted that the fractionation reactions initially proposed by \cite{terzieva2000} work no better in this respect. Accuracy in observational measurements of nitrogen isotopic ratios, in \ce{N2H+} and/or other species, has also improved significantly, suggesting that the discrepancy between observations and models could be due to an incomplete understanding of nitrogen fractionation, or even mistaken concepts of the nitrogen chemistry.

The main originality of the present study is to present time-dependent calculations of the isotopic ratio of nitrogen-bearing species during the early phases of the gravitational collapse of a {pre-stellar} core. Indeed, all previous studies \citep{terzieva2000, charnley2002, rodgers2008a, hilyblant2013b, roueff2015, wirstrom2018, loison2019} have considered constant density models while focusing on the chemical aspects. Here, we consider a collapsing core starting from steady-state initial conditions, and explore various chemical processes with the hope of finding potential solutions to the problems outlined above. The following Section~\ref{sec:model} summarizes the salient characteristics of the collapse model. In \rsec{results}, we compare the computed degrees of fractionation of nitrogen with the results of previous calculations and with observations. In \rsec{discussion}, the obtained results are discussed and compared to previous studies. Our conclusions are given in \rsec{conclusions}.

\section{The model}
\label{sec:model}

\subsection{Physical and chemical model}
\label{sec:physmodel}

The model used in this study was first presented in \cite{hilyblant2018a} (hereafter \hba). Detailed information on the properties and validation of the model are given in \hba\ and we here summarize its central assumptions and features. Our model follows the evolution of the chemical composition of a nonrotating, spherical cloud of gas and dust undergoing isothermal collapse. The chemistry is followed self-consistently with the collapse, whose treatment was inspired by the results of \cite{larson1969} and \cite{penston1969}: The cloud has a free-falling core of {radius $R_c$ and} uniform, but increasing, density \rhocore, which is surrounded by an envelope with a density profile given by $\rhoenv(R)=\rhocore(R/R_c)^{-2}$. As the collapse proceeds, the mass of the inner core decreases whilst that of the envelope increases by advection of gas and dust across the interface with the core, such that the total mass of the cloud $M = 4\pi R_0^3 \rho _0/3$
%\begin{equation}
%	M = \frac {4\pi }{3} R_0^3 \rho _0
%\end{equation}
remains constant. In this expression, the initial external radius is denoted by $R_0$ and the initial mass density by $\rho _0$.

{The simple model above captures quite well the Larson-Penston (L-P) solution for the time-dependent density during the early stages of collapse, before the formation of the first hydrostatic (Larson) core at a typical density $\nh= n(\rm H) + 2n(\hh) \sim10^{10}$\ccc\ ($\rhocore\sim10^{-14}$g\ccc) \citep{larson1969, young2019} beyond which our model loses validity. In what follows, we study the initial phase of collapse and focus on the evolution of chemical abundances up to the stage corresponding to a central density no higher than \dix{8}\ccc. The central density, which is a monotonically increasing function of time, parametrizes the evolution of abundances.  We note also that the model allows us to compute separately the contributions of the envelope and of the inner core to the total column density of any species.}

The initial chemical conditions are derived assuming that the cloud had time to reach a steady state prior to collapse. The subsequent chemical evolution of the 3-fluid medium, which comprises neutral, positively and negatively charged species, is followed by means of conservation equations of the form
\begin{equation}
\frac {1}{n} \frac {{\rm d}n}{{\rm d}R} = \frac {N}{n v} - \frac {2}{R},
\label{equ1}
\end{equation}
where the number density of the species is denoted $n(R)$ [\ccc], and $N(R)$ [\ccc\pers] is its rate of formation per unit volume through chemical reactions; the flow speed is given by $v(R) = {\rm d}R/{\rm d}t$. Grain coagulation is followed self-consistently during the collapse \citep{flower2005}. The mass density of each of the three fluids is determined by the corresponding conservation equation
\begin{equation}
	\frac {1}{\rhoenv} \frac {{\rm d}\rhoenv}{{\rm d}R} = \frac {S}{\rhoenv v} - \frac {2}{R},
	\label{equ2}
\end{equation}
where $\rhoenv (R)$ [g\ccc] denotes the mass density, and $S(R)$ [g\ccc\pers] is the rate of production of mass per unit volume. Throughout the paper, abundances are expressed relative to hydrogen nuclei, $n(\ce{X})/\nh$, and the isotopic ratio of any species \ce{X}, noted \rx{X}, may refer to abundance or column density.

\subsection{Chemical network}
\label{sec:chemnet}
%\subsubsection{\fifn\ chemistry}
For the purpose of the present study, the condensed\footnote{The UGAN network exists in a condensed version and one in which nuclear symmetry of all hydrides are treated explicitly.} UGAN network (see \hba) has been extended to include the $^{15}$N isotopologs {(see \rapp{network}).}
%assuming identical branching ratios for the corresponding output channels, and has been limited to singly-substituted species. The following reactions were treated manually assuming that they proceed by proton-hop and deuteron-hop:
%\begin{align}
%%\ce{N2H+ +  NH3  -> NH4+ + N2} \label{equ3}&&\\
%%\ce{N2D+ + NH3 -> NH3D+ + N2}\label{equ4}\\
%%\ce{N2H+ + NO -> HNO+ + N2}\label{equ5}
%\cee{
%	&N2H+ +  NH3 -> NH4+ + N2 \\
%	&N2D+ + NH3 -> NH3D+ + N2\\
%	&N2H+ + NO -> HNO+ + N2}
%\end{align}
A ``sanity'' check of the duplicated network was undertaken to verify that the $^{14}$N and $^{15}$N variants are independently conserved when no fractionation reactions are included.

We have considered two sets of fractionation reactions and rate coefficients, compiled by \cite{terzieva2000} and \cite{roueff2015} (hereafter TH00 and R15, respectively{; see Table~\ref{tab:frac}}), with a two-fold objective: First, to enable a direct comparison with our previous study (\hba); and second, to compare the dependence of our results on the two sets available in the literature. Our fiducial network does not take into account the possibility that \ce{CN + N} could lead to isotopic exchange \citep[see also ][]{wirstrom2018}. The modifications to  \citetalias{roueff2015} made by \cite{loison2019} have also been tested but will not be considered in detail since they were found to have a negligible effect on the isotopic ratios.

The bi-molecular reaction rate coefficients in our network are expressed as
\begin{equation}
	\label{eq:rate}
	k(T) = \alpha (T/300)^\beta \exp(-\gamma/k_{\rm B}T),
\end{equation}
and hence the form of the rate coefficient for the N$^+$ + N$_2$ fractionation reaction, adopted  by  \citetalias{roueff2015}, could not be used. Nevertheless, because the kinetic temperature in our models is assumed fixed and equal to 10~K, our simplified expression for the reverse reaction, $k(T)=2.4\times 10^{-10}\exp(-28.3/T)$, deviates from theirs by only 3\% at $T = 10$~K, and by less than 20\% for $T \le 30$~K. The rate of the critical reaction \ce{N+ + H2 -> NH+ + H} was approximated to the rate with o-\hh\ \citep{dislaire2012}, {which is appropriate for the ortho-to-para ratio of \hh\ ($\sim\dix{-3}$) and the low kinetic temperature considered here}. The rate for the deuterated variant is taken from \cite{marquette1988}.

Since \cite{hilyblant2013b}, new calculations of the rates of dissociation and ionization by cosmic-ray induced secondary photons have become available \citep{heays2017}, and the results of these calculations have been incorporated in the present model (see also \hba).
{Furthermore, compared to \hba, cosmic-ray induced desorption of water and ammonia has been revised by \cite{faure2019}. As a result, desorption of ammonia (and water) via direct impact of cosmic-rays is found to be non-negligible even at 10~K owing to the prompt desorption mechanism \citep{bringa2004}. In this context, we emphasize that grain-surface chemistry in the UGAN network is handled in a simplified way, where sticking atoms are assumed to react with hydrogen on the surface, forming saturated molecules such as \ce{CH4}, \ce{H2O}, and \ce{NH3} (see \rtab{grain_ads}). The binding energies of the \foun\ and \fifn\ variants of adsorbed species have equal values.}

The initial (deuterated) UGAN network consists of 1196 reactions linking 156 species\footnote{Available upon request to the corresponding author.}. After including the \fifn\ variants of nitrogen-bearing species and the fractionation reactions, the \fifn\ version of UGAN contains $\approx 1720$ reactions and 205 species.

\subsection{Model parameters and initial composition}
\label{sec:setup}

\begin{table}
	\caption{\label{tab:abinit}Partition of the elemental abundances$^\dagger$ in the gas phase and in the grain mantles, as adopted when specifying the initial composition of the gas, prior to its evolution to chemical steady state.}
	\label{abelem}
	\centering
	\begin{tabular}{lcccc}
		\toprule
		Element & Total & \mc{2}{Volatile} & Refractory \\
		& & Gas & Mantles$^\S$ \\
		\midrule
		H   &   1.0       & 1.0       &           &            \\
		D	& 1.60(-5) \\
		He	&   1.00(-1)  & 1.00(-1)  &           &    \\
		C	&   3.55(-4)  & 8.27(-5)  & 5.55(-5)  & 2.17(-4)\\
		N	&   7.94(-5)  & 6.39(-5)  & 1.55(-5)  &           \\
		$^{15}$N$^\ddagger$
		& 2.41(-7) & 1.94(-7) & 4.70(-8)  \\
		O	&   4.42(-4)  & 1.24(-4)  & 1.78(-4)  & 1.40(-4)\\
		S	&   1.86(-5)  & 6.00(-7)  & 1.82(-5)  &            \\
		Mg  &   3.70(-5)  &           &           & 3.70(-5)   \\
		Si  &   3.37(-5)  &           &           & 3.37(-5)   \\
		Fe  &   3.23(-5)  & 1.50(-9)  &           & 3.23(-5)\\
		\bottomrule
	\end{tabular}
	\tabnotes Numbers in parentheses are powers of 10. $\dagger$~Abundances are expressed relative to the hydrogen nuclei density, \nh. $\S$ The initial composition of the mantles can be found in Table 3 of \hba. $\ddagger$ The isotopic ratio of {bulk} nitrogen is the {adopted} present-day value \rtot=330.
\end{table}

As in \hba, our model calculations of the evolution of atomic and molecular abundances proceed in two steps: First, a calculation at constant density {(hereafter step 1)} is followed until a steady state is attained. In this first step, gas-grain processes (adsorption and desorption) {and grain-surface chemistry} are ignored. The radius of the grains includes the contributions of a refractory core and of an initial ice mantle; it remains constant and intervenes only in establishing the free-electron concentration, through electron attachment. Second, the abundances are computed following our L-P prescription of the collapse {(hereafter step 2)}. The initial abundances are the steady-state values, and adsorption and desorption processes {and grain-surface chemsitry} are included.
%We adopt the L-P simulation, rather than assuming free-fall; but our previous study \citep{hilyblant2018a} demonstrated that both models lead to similar results and conclusions.

In the present calculations, the rate of cosmic-ray ionization of \hh\ was $\zeta_\hh = 3 \tdix{-17}$\pers. The refractory core has a radius \acore=0.1~\micr\ and a mass density $\rhogr=2\gccc$, while that of the mantles is $\rho_m=1\gccc$. The total grain radius, including the mantles, is thus initially $\agr=0.128$~\micr\ (see \req{agr}). The constant density models have \nh=\dix{4}\ccc, and the kinetic temperature is 10~K in both steps. As in \hba, the total mass $M$ is equal to the Jeans mass of a cloud with density \nh=\dix{4}\ccc\ and temperature \tkin=10~K, viz. $M=M_{\rm J}\approx 7\msol$, which is also close to the critical mass of a pressure-truncated, isothermal, Bonnor-Ebert sphere. The free-fall timescale of such a sphere of gas is $\tauff = (3\pi/32 G \rho_0)^{1/2}=4.4\tdix{5}$~yr.

\section{Results}
\label{sec:results}

\subsection{CO depletion and the timescale of collapse}
\label{sec:co}

\begin{figure}
	\centering
	\includegraphics[width=0.8\hsize]{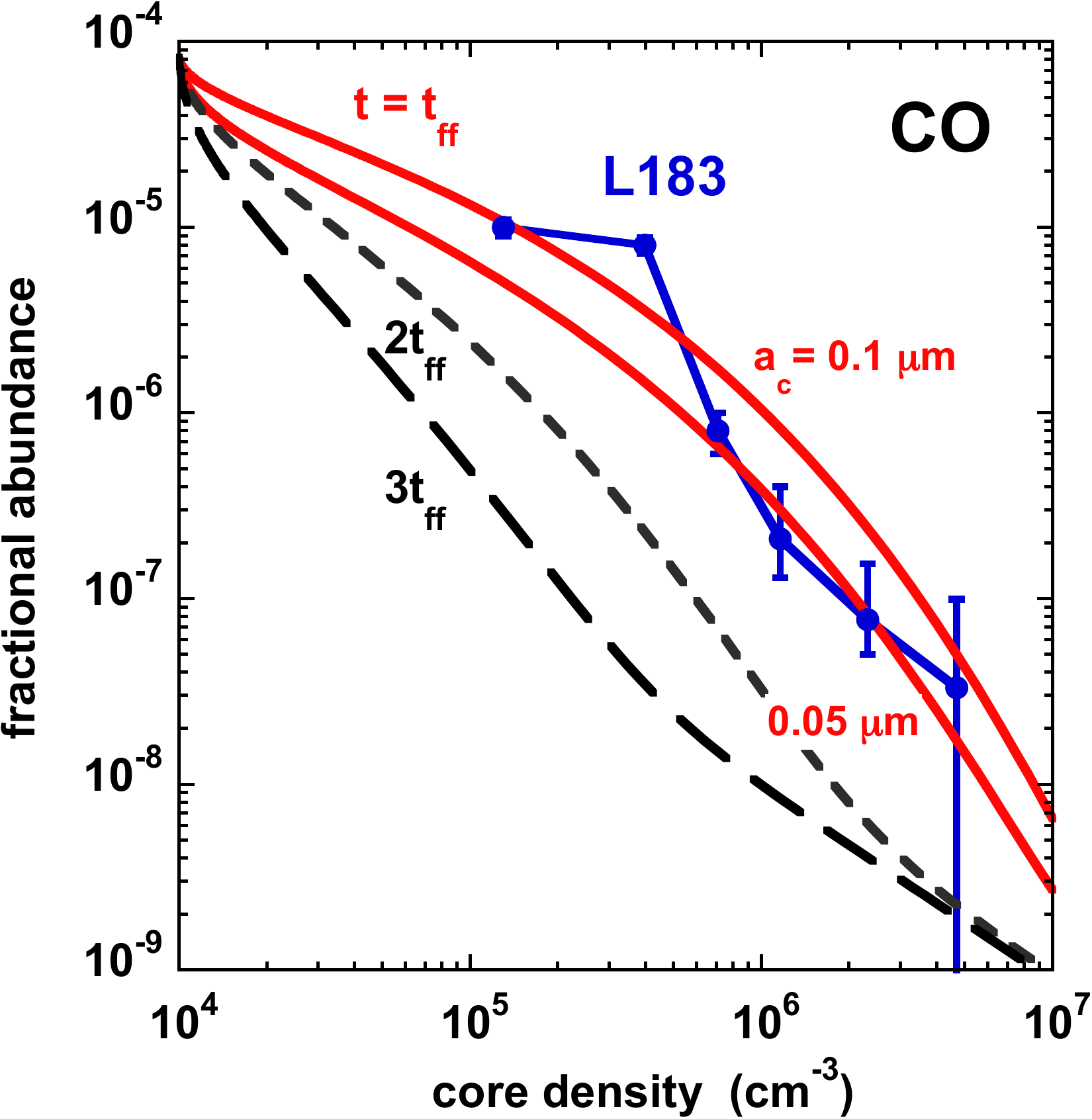}
	\caption{The depletion of CO with increasing density, as predicted by our L-P collapse models for three values of the dynamical timescale, $t_{\rm dyn}$, and $\acore = 0.10$~\micr, compared with the observational results of {\protect\cite{pagani2007, pagani2012}} for L183. The L-P calculations corresponding to $t_{\rm dyn} = t_{\rm ff}$, where $t_{\rm ff}$ is the free-fall time, are shown also for $\acore = 0.05$~\micr. }
	\label{fig:COdepletion}
\end{figure}

The time required for nitrogen-bearing species to reach steady state (often called chemical equilibrium), through gas-phase reactions can exceed 1~Myr for a cloud of density $\nh =10^4$\ccc. Since the quasi-static evolution of cores may last 0.5--1~Myr \citep{brunken2014}, it is not guaranteed that the initial abundances of these species in the gas phase, prior to collapse, are equal to their values in steady state. Nevertheless, with the possible exception of the role played by the N$_2$:N abundance ratio in the nitrogen chemistry, the abundances when collapse sets in are not crucial to the calculation of the composition of the gas during its subsequent evolution providing that its composition adapts rapidly to the prevailing physical conditions. As the cloud collapses, and its density increases, this condition will tend to be satisfied. However, freeze-out on to the grains occurs simultaneously, reducing the abundances of molecules in the gas phase (and hence the likelihood of their being observed). It follows that the assumption of an initial chemical equilibrium may influence the predicted composition of the gas during the early stages of its gravitational collapse; this caveat should be borne in mind when considering the results in Section~\ref{sec:collapse}.

%\footnote{This also assumes that the chemical system is far from critical points of chemical bistability \citep{pineau1992, lebourlot1993}.}

A key question in star-formation studies is the dynamical timescale, which may be constrained by time-dependent models of collapse that include the gas-grain chemistry, since both gas-phase and adsorption processes depend on the density. In our nonrotating, magnetic-pressure-free model, the dynamical timescale is that of the gravitational collapse, \tauff. Nevertheless, the collapse may be slowed by magnetic fields and proceed on the ambipolar diffusion timescale, which can be up to 10 times larger than $\tau_{\rm ff}$. Such a delay can be introduced in our model calculations by increasing artificially the dynamical timescale \citep{flower2005}. However, delaying the collapse enhances freeze-out which, if not counterbalanced by desorption, will lead to lower gas-phase abundances.

In Fig.~\ref{fig:COdepletion}, we compare the depletion of CO with increasing density\footnote{The density to which reference is made in the models of gravitational collapse is that of the core (cf. \rsec{physmodel}).}, as predicted by our L-P simulation, with the observations of L183 by \cite{pagani2007, pagani2012}. Results are shown for collapse timescales corresponding to 1, 2, and 3 times $\tau_{\rm ff}$ and an initial grain-core radius $\acore$=0.1\micr. It may be seen that, when the dynamical timescale is equal to the free-fall time, there is striking agreement with the observational results of \cite{pagani2007, pagani2012}. On the other hand, when the timescale for collapse is increased, the agreement with the observations deteriorates significantly. Short collapse timescales, $\le 0.7$~Myr, are also supported by an independent observational constraint, based on the D/H ratio in \ce{N2H+} \citep{pagani2013}. Also displayed in this Figure is the result of calculations with $\acore$=0.05\micr\ for a dynamical timescale equal to $\tau_{\rm ff}$, showing that the agreement with the observed abundances is preserved.

In what follows. we thus consider cores collapsing on a free-fall timescale.

\subsection{Steady-state abundances and isotopic ratios}
\label{sec:abss}
%\phb{To be shortened}

\begin{figure}
%	\centering
%	\includegraphics[height=.3\hsize]{SS_TH00_ab.png}
%	\hfill
%	\includegraphics[height=.3\hsize]{SS_TH00_ratios.png}
%	\hfill
%	\includegraphics[height=0.3\hsize]{SS_R15_ratios.png}\\
%	\includegraphics[height=0.3\hsize]{LP_TH00_ab.png}
%	\hfill
%	\includegraphics[height=0.3\hsize]{LP_TH00_ratios.png}
%	\hfill
%	\includegraphics[height=0.3\hsize]{LP_R15_ratios.png}
	\centering
	\includegraphics[width=\hsize]{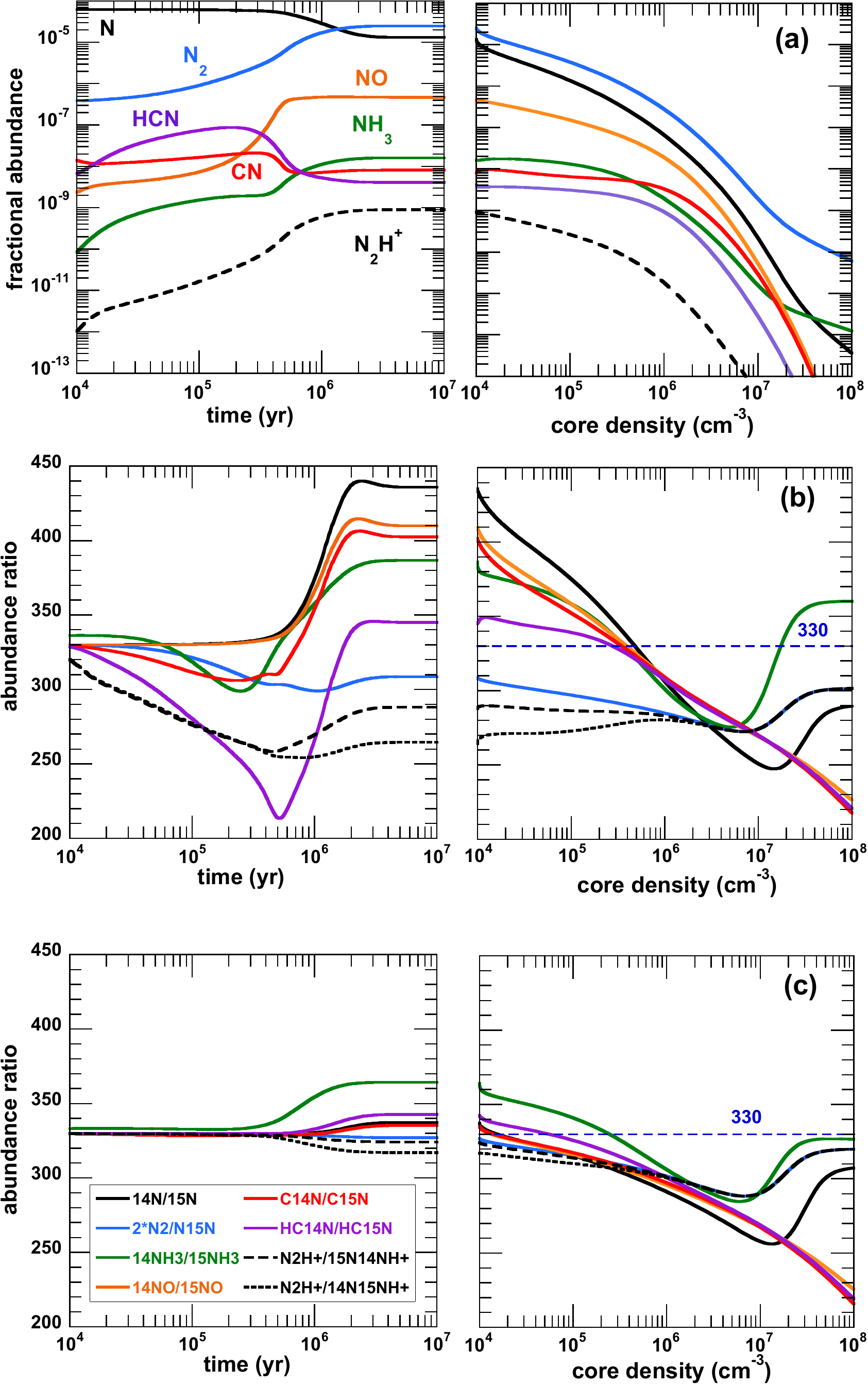}
	\caption{Results from chemical models. In each row, the left panel shows the temporal evolution during the quasi-static evolution of the core {(step 1)}, and the right panel shows the evolution of the inner core as a function of the central density during the collapse {(step 2)}. The  {bulk \foun:\fifn} isotopic  ratio is \rtot=330. \textit{(a)} Gas-phase abundances, relative to \nh, of the main isotopologs of relevant species. \textit{(b)} \rr\ abundance ratios when adopting the \citetalias{terzieva2000} set of fractionation reactions. \textit{(c)} Same as \textit{(b)} but with the fractionation reactions from \citetalias{roueff2015}.}
	\label{fig:ss_lp}
\end{figure}

\begin{table}
	\centering
	\caption{Steady--state abundances and $^{14}$N:$^{15}$N isotopic ratios, as calculated for our reference model (see \rsec{setup}). The {bulk} isotopic ratio $^{14}$N:$^{15}$N is \rtot=330.}
	\label{tab:abss}
	\begin{tabular}{llll}
		\toprule
		Species    & Abundance$^\dagger$ & 
		\citetalias{terzieva2000} & \citetalias{roueff2015} \\
		\midrule
		N          & 1.3(-05)  & 436                       & 337                     \\
		N$_2$      & 2.5(-05)  & 309                       & 327                     \\
		NH$_3$     & 1.6(-08)  & 387                       & 364                     \\
		CN         & 8.2(-09)  & 402                       & 335                     \\
		NO         & 4.7(-07)  & 410                       & 335                     \\
		HCN        & 4.1(-09)  & 345                       & 343                     \\
		N$_2$H$^+$ & 9.1(-10)  & 288$^{(1)}$               & 324$^{(1)}$             \\
		N$_2$H$^+$ & 9.1(-10)  & 265$^{(2)}$               & 317$^{(2)}$             \\
		\hline
	\end{tabular}
	\tabnotes Numbers in parentheses are powers of 10. $\dagger$ Expressed relative to $n_{\rm H}$. $^{(1)}$ \nnhp/\qnnhp. $^{(2)}$ \nnhp/\nqnhp.
\end{table}

In the left panels of Fig.~\ref{fig:ss_lp} are plotted the evolution toward steady state of the gas-phase fractional abundances of nitrogen-containing species {during step 1}, and the predicted isotopic ratios using the reactions and rate coefficients of either \citetalias{terzieva2000} or \citetalias{roueff2015}. The initial gas- and ice-phase abundances of the elements are listed in Table~\ref{tab:abinit}, with hydrogen in molecular form and metals in atomic form, either neutral (\foun, \fifn, O) or singly ionized (\ce{C+}, \ce{S+}, \ce{Fe+}).

Our calculations agree with previous studies \citep{wirstrom2018} in that the degrees of fractionation are globally low; this is in agreement with the directly measured ratios in starless cores and star-forming regions summarized in Table~\ref{tab:review}. The degrees of fractionation are lower when the reaction rates of \citetalias{roueff2015} are adopted, rather than those of \citetalias{terzieva2000}. We note significant differences between the present calculations and those of our previous study \citep{hilyblant2013b}, which are due primarily to the update of the rates of dissociation and ionization by the {cosmic-ray induced} ultraviolet radiation field, following the results of \cite{heays2017}. For example, the rate of dissociation of N$_2$ is smaller by a factor of approximately 25 than the rate used previously, which explains the higher ratio of the steady-state abundances of N$_2$ and N, $n({\rm N}_2)/n({\rm N})$, in the present case.

We have investigated the effects of varying ill-defined parameters such as the grain-core radius, \acore, $\zeta_\hh$, the elemental S abundance, and the C:O elemental abundance ratio (in practice, the carbon abundance). The effects on the degrees of fractionation of $^{15}$N were generally modest, being somewhat more pronounced when the rate coefficients of \citetalias{terzieva2000} were adopted, rather than those of \citetalias{roueff2015}.

\subsection{Fractionation during gravitational collapse}
\label{sec:collapse}

The principal question that we wish to answer is whether the degrees of fractionation of the nitrogen-bearing species, computed in steady state, vary significantly in a subsequent gravitational collapse. Our setup is described in \rsec{setup}.

As the density of the medium increases, molecules ultimately freeze on to the grains, building up layers of ice that contribute, in addition to grain coagulation, to the grains' overall radius. Cosmic-ray induced processes restore molecules to the gas phase from the grain mantles, but the rates of evaporation increase less rapidly than the rates of freeze-out, and hence molecular species become depleted from the gas phase.

The evolution of the abundances and \foun:\fifn\ abundance ratios with the central density (or, equivalently, time) {during step 2}, for selected species, is shown in the right panels of Fig.~\ref{fig:ss_lp}. As can be seen, the fractional abundances decrease with increasing density, for all species. However, not all species behave identically as the density increases, and the differences among species are more pronounced with the \citetalias{terzieva2000} reaction rates. In this case, the isotopic ratio in atomic nitrogen decreases from its steady-state value of 440 to 246 when the density has reached \nh=\dix{7}\ccc. Other species, directly related to atomic nitrogen, namely CN and NO, also see their isotopic ratios decrease by significant factors. To a lesser extent, this is also true for species, such as \ce{NH3} and HCN, that are more indirectly related to N. In contrast, the isotopic ratios of \ce{N2} and \ce{N2H+} are approximately constant during the collapse. When adopting the  \citetalias{roueff2015} rates, the variations are obviously smaller because the steady-state ratios are closer to the isotopic ratio of {bulk} nitrogen. Nevertheless, there are variations among species.

We find that, with both the \citetalias{terzieva2000} and \citetalias{roueff2015} fractionation reactions, the gas becomes gradually enriched in the heavier isotope of ammonia in the early stages of the collapse, but this tendency ultimately reverses as desorption of the initial reservoirs\footnote{\cite{smith2015} appealed to the significance of ice reservoirs and gas-grain interactions when attempting to explain their high observed values of the gas-phase $^{12}$CO:$^{13}$CO in solar-type Young Stellar Objects.} of the $^{14}$NH$_3$ and $^{15}$NH$_3$ ices begins to be significant (see Fig.~\ref{fig:ss_lp}). We recall that $^{14}$NH$_3$ and $^{15}$NH$_3$ ices are assumed to be initially present in the grain mantles in an isotopic ratio of 330. At high density, ammonia experiences an increase in its isotopic ratio, which is also observed for both isotopologs of \ce{N2H+}.

It may be noted that, in the course of the collapse, the gas-phase fractionation reactions have little influence, relative to adsorption to and desorption from the grains. Consequently, the differences in the degrees of fractionation, as calculated with the fractionation reactions of  \citetalias{terzieva2000} and \citetalias{roueff2015}, are attributable essentially to the differences in their initial, steady-state values.

\section{Discussion}
\label{sec:discussion}

\subsection{Depletion-driven fractionation}
\label{sec:depletion}
%f\revised{12}

%\begin{figure}
%  \centering
%  \includegraphics[width=0.95\hsize]{depletion_ratios_publi_a.png}
%  \includegraphics[width=0.95\hsize]{depletion_ratios_publi_b.png}
%  \caption{Evolution, for a constant density model, of the isotopic ratios N, \ce{N2}, HCN, and \ce{NH3} normalized to the bulk ratio of 330 (top panel), and of their abundances (bottom panel). The fractionation reaction set is that of Roueff et al 2015. The full curves are for models with adsorption only, while dashed curves are for models with both adsorption and desorption. In the bottom panel, desorption was not included. The evolution of the dust grain radius $a_g$ and of the freeze-out timescale, \taufo, for N (lower curve) and \ce{N2} are also shown.}
%  \label{fig:depletion}
%\end{figure}

The decrease in the isotopic ratios with increasing density during collapse seen in \rfig{ss_lp} {(step 2, right-hand panels)} is due to adsorption: Owing to their higher mass, $^{15}$N-containing species have slightly slower thermal speeds than their $^{14}$N variants, and hence their collision frequencies with the grains are lower \citep[see also][]{loison2019}. The later increase in the ratios at densities above $\sim$5\tdix{6}\ccc\ is instead a result of desorption by cosmic-rays and secondary photons.

One key parameter in the adsorption-driven fractionation is the freeze-out timescale, which can be written $\taufo = x_0 \mamu^{1/2}$ (see \req{taufo_x0}) with $\mamu$ the atomic mass number and $x_0$ a factor which depends on the density, temperature and on the grain properties (size, mass density). For the grain parameters adopted in this study (see \rsec{setup}), one obtains {$x_0 = 7.9\tdix{4}\rm\,yr$}. In a constant density, isothermal, model, and ignoring adsorption/desorption processes $x_0$ is constant: The difference in adsorption rates for the \foun\ and \fifn\ variants is thus $\delta\taufo/\taufo = 1/2 (\delta \mamu/\mamu)$ or 1.8\% for HCN, and twice larger for N atoms. It follows that the isotopic ratio \rrr\  would decrease with time as
\begin{equation}
	\rrr = \rrr_0 e^{-t/\taufr},
	\label{eq:depletion}
\end{equation}
where $\rrr_0$ is the initial isotopic ratio and the $e$-folding timescale is given by (see \req{taufr})
\begin{equation}
\taufr\approx 2 x_0 \mamu^{3/2}\approx 2 \mamu \taufo.
\label{eq:tauR}
\end{equation}
The depletion-driven fractionation timescale, \taufr, is thus significantly longer than the freeze-out one.

Our simple analytical estimate agrees well with the result of the constant-density study of nitrogen fractionation by \cite{loison2019} for HCN, but not for HNC. Furthermore, these authors report variations by several tens of percent for the isotopic ratio of atomic nitrogen after typically 0.4 Myr, while the above derivation predicts a decrease of 13\% only (see Appendix \rapp{depletion}). The above description of depletion-driven fractionation is also in good agreement with our constant-density calculations, shown in \rfig{depletion}. For our model parameters, and with \agr=0.128~\micr, the timescale for depletion-driven fractionation is $\taufr=$8.3~Myr for N atoms, and 22~Myr for HCN molecules.

From Eq.~\ref{eq:tauR} one would expect to see a dependence of $\taufr$ upon the mass of the depleting species, which is not observed in the models. Instead, all species follow a single exponential decay which coincides with that of atomic nitrogen. This result indicates that the temporal variation of \rrr(t) is determined primarily by the differential depletion of \foun\ and \fifn, which propagates to other species through gas-phase chemistry. It may be verified that the timescale for the formation of N-bearing gas-phase species ($\sim 4$~Myr at a time $t=0.1$~Myr and 1~Myr at $t=0.3$~Myr) is significantly longer than the freeze-out timescale of N atoms but shorter than $\taufr$ for the N atoms, and much shorter than $\taufr$ for the heavy species ($m\approx 28$ amu). Therefore, the freeze-out of N-bearing molecules will reflect \rrr(t) of atomic N. Eventually, desorption brings the gas-phase isotopic ratios back toward the value of 330 (see \rfig{depletion}) that is adopted for the initial composition of the ices.

When collapse is taken into account, the density increases and hence the freeze-out timescale $\taufo\sim1/\nh$ decreases faster than the free-fall one, $t_{\rm ff}\sim \nh^{-0.5}$. The effect of differential depletion is thus more pronounced than in a constant density model. Indeed, as \rfig{ss_lp} shows, \rrr\ decreases by more than 10\%, when using the  \citetalias{roueff2015} rates, for most N-bearing molecules as the density increases from \dix{4} to \dix{6}\ccc. {As the density increases above $\sim5\tdix{6}\ccc$, desorption by direct cosmic ray impact liberates \fifn-poor ammonia ices in the gas-phase thus raising the value of \rrr(\ce{NH3}). The same mechanism also operates for other species but at higher densities (not shown in \rfig{ss_lp}) The decrease in \rrr\ due to differential depletion is significantly larger---reaching $\sim$30\% for atomic nitrogen, NO, and CN---when the  \citetalias{terzieva2000} rates are used. Calculations at higher temperatures show that, in a warmer gas, desorption would occur earlier during collapse, at densities typically $\sim10$ times smaller at a temperature \tkin=30~K.}

\subsection{Comparison with observations}

\begin{figure}
	\centering
	\includegraphics[width=0.48\hsize]{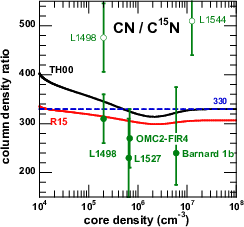}\hfill%
	\includegraphics[width=0.48\hsize]{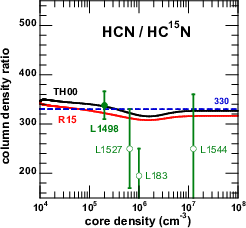}\bigskip\\
	\includegraphics[width=0.48\hsize]{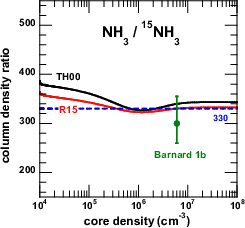}\hfill%
	\includegraphics[width=0.48\hsize]{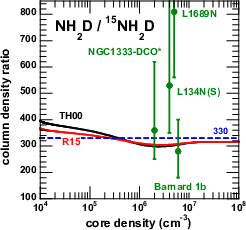}\\
	\caption{The column density ratios of $^{14}$N- and $^{15}$N-containing species in the course of L-P collapse {(step 2)}, compared with the values observed in pre-stellar cores through direct (filled symbols) and indirect (open symbols) measurements (see \rfig{obs} and \rapp{review}).}
	\label{fig:obs}
\end{figure}

In \rfig{obs}, we compare the \foun:\fifn\ column density ratios of several nitrogen-containing species to the values obtained through direct measurements in pre-stellar cores (see \rapp{obs}). The case of \ce{N2H+} is shown in Fig.~\ref{fig:n2hp} and is considered further in \rsec{n2hp}.
%, the overall level of agreement with the observations appears better for some species than others. The agreement owes principally to the increase in the degrees of enrichment in $^{15}$N during the early stages of gravitational collapse.

\subsubsection{Nitriles}

Because direct measurements in CN and HCN yield essentially non-fractionated values of \rrr\ (see \rtab{review}), it is not surprising that our model predictions are in relatively good agreement with the observed ratios \citep[see][]{wirstrom2018}. However, the new result here is that this remains true during the collapse, which brings the gas-phase isotopic ratios back toward the value of 330.

\rfig{obs} appears to indicate that the direct measurement of \rx{CN} toward L1498---which has been re-analyzed using a non-LTE model of the radiative transfer in spherical geometry (see \rapp{cn})---favors the reaction set of  \citetalias{roueff2015} to that of  \citetalias{terzieva2000}. The ratio toward Barnard~1b is compatible with both reaction sets, although it suggests that \rx{CN} could be lower than 330. Regarding HCN, the only direct measurement was obtained toward L1498, and our model predictions, using either  \citetalias{terzieva2000} or  \citetalias{roueff2015}, are in agreement with the observed value. The indirectly determined ratio toward L183 is significantly lower than the model predictions and the isotopic ratio for {the bulk} nitrogen. We note that it is impossible to anticipate the sense in which this ratio would change if a detailed radiative-transfer model were to be used in the analysis of the observations; the same holds true for the indirect measurement of \rx{CN} toward L1544.

%In the case of the early-stage core L1498, both CN and HCN are well reproduced by either the TH00 or  \citetalias{roueff2015} networks. The HCN ratio in the L1544 core, with higher a central density, is also well matched by our collapse model. The CN ratio is largely underestimated, in contrast with L1498, which may be due to the reported observed ratio being an LTE value, while the CN ratio in L1498 was obtained with a detailed 1D, non-LTE, model (see \rsec{app_cn}).

\subsubsection{\ce{N2H+}}
%\revised{12}
%\phb{@ALEX: a revoir}
\label{sec:n2hp}

\begin{figure}
	\centering
	\includegraphics[width=0.7\hsize]{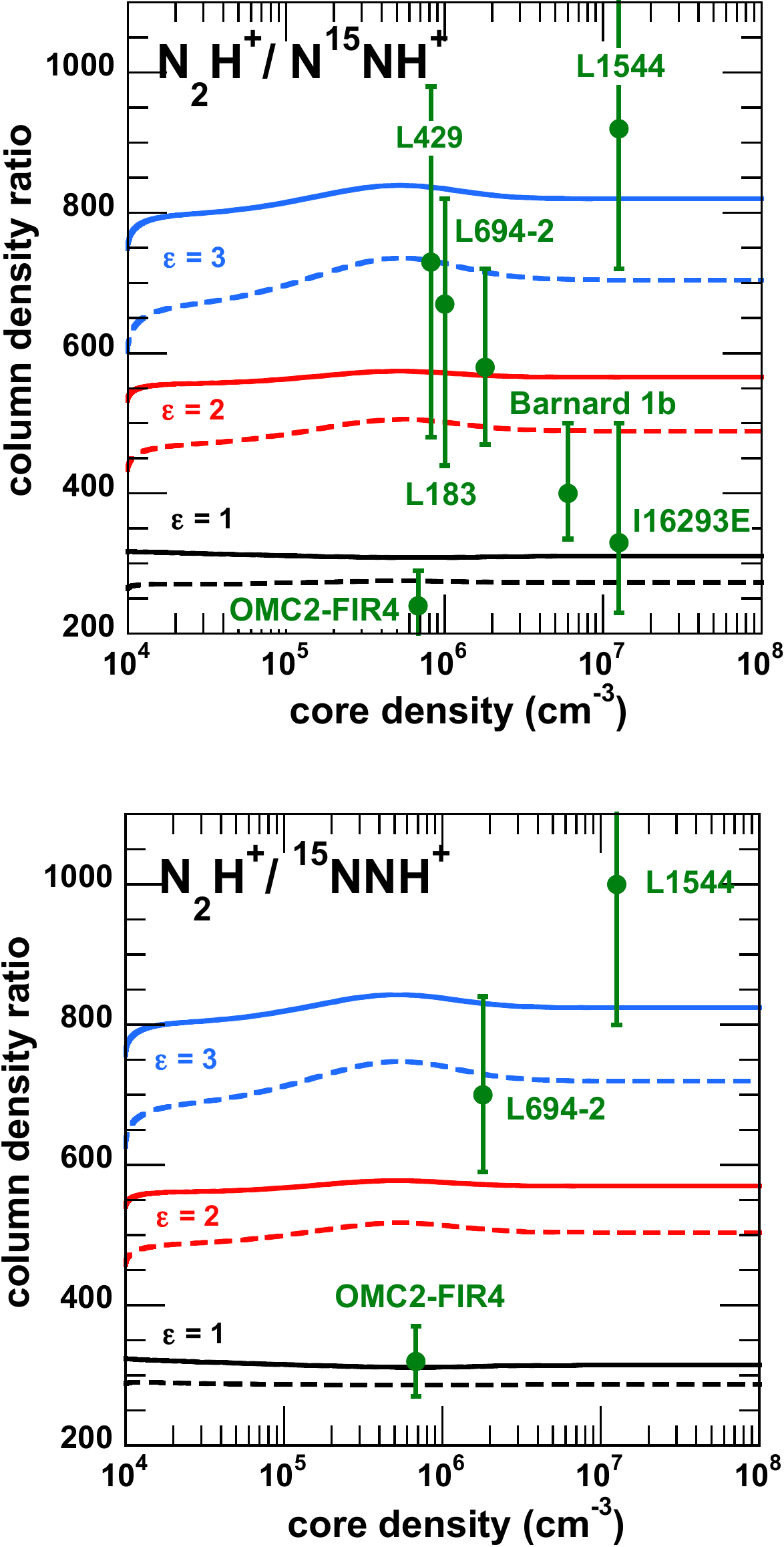}
	\caption{The isotopic ratio (column density) of \ce{N2H+} as a function of the core density (\nh) in the course of L-P collapse {(step 2)}, compared with the values observed in pre-stellar cores (see also \rtab{review}). The abundance ratios N$_{2}$H$^+$:N$^{15}$NH$^+$ and N$_{2}$H$^+$:\ce{$^{15}$NNH+} are shown for three values of the enhancement factor, $\epsilon$, of the rate coefficient for dissociative recombination of N$^{15}$NH$^+$ (see \rsec{n2hp}), viz. $\epsilon = 1$, 2, and 3. Results with rates from both  \citetalias{roueff2015} (full curves) and  \citetalias{terzieva2000} (dashed) are shown.}
	\label{fig:n2hp}
\end{figure}

The results of our fiducial model for \ce{N2H+} are shown as black lines in \rfig{n2hp}. In practice, our results fit the observations no better than previous attempts, by other groups. Thus, this species provides evidence of an unresolved issue with the nitrogen chemistry.

\qnnhp\ and \nqnhp\ form mainly in the proton-transfer reaction of $^{15}$NN with H$_3^+$, for which the rate coefficient has been taken identical to that for the reaction N$_2$(H$_3^+$, N$_2$H$^+$)H$_2$. Were the rate coefficients of the reactions $^{15}$NN(H$_3^+$, $^{15}$NNH$^+$)H$_2$ and $^{15}$NN(H$_3^+$, N$^{15}$NH$^+$)H$_2$ to be significantly smaller, the depletion of the $^{15}$N-bearing isotopic forms could be understood. However, the rate coefficients for these barrierless exothermic proton-transfer reactions are presumably independent of the isotopic form of nitrogen.

Alternatively, enhanced rates of dissociative recombination (DR) of $^{15}$NNH$^+$ and N$^{15}$NH$^+$ with electrons at $T$=10~K---the main destruction mechanism---might account for their depletion, as proposed also by \cite{loison2019}. While the electronic potential curves are identical for all isotopologs, the vibrational and rotational levels are shifted due to the differing masses, changing their position relative to the dissociation curve. The measurements of \cite{lawson2011} have shown that, at 300~K, the DR of N$_2$H$^+$ proceeds faster than that of $^{15}$N$_2$H$^+$, by about 20\%. A larger effect, in the opposite sense, is possible at very low temperature, where an indirect mechanism dominates in the recombination of N$_2$H$^+$ \citep{dossantos2016}. This tentative explanation requires experimental or theoretical confirmation, and we note in this context that calculations are in progress, using Multichannel Quantum Defect Theory (Dahbia Talbi, private communication).

%To test quantitatively this effect, we have considered models where the DR rate of \nnhp\ is enhanced by a factor $\epsilon$ compared to the DR rate of its \fifn-variants.  
Our model confirms the increase in the isotopic ratio of \ce{N2H+} when the DR rate of \qnnhp\ and \nqnhp\ is multiplied by a factor of $\epsilon$ compared to the DR rate of \nnhp. \rfig{n2hp} shows that the isotopic ratio scales approximately linearly with $\epsilon$; this is particularly true at high densities: The ratio rises from 330 to 560 for $\epsilon=2$, and reaches $\approx 825$ when $\epsilon=3$. Thus, it appears that a factor of two to three enhancement in the DR rate coefficients yields values of \rrr\ for both \nqnhp\ and \qnnhp\ which are compatible with those observed by \cite{redaelli2018}. We also note that the enhancement preserves the slightly lower value of the ratio for \ce{N2H+}/\nqnhp\  compared to \nnhp/\qnnhp\ \citep{kahane2018, redaelli2018}.

{In \rfig{review}, the isotopic ratios in \ce{N2H+} are shown with the sources sorted by increasing temperature from left to right, with L1544 and L183 being the coldest and OMC2-FIR4 the warmest. Although there are uncertainties in the temperature determination, I16293E pre-stellar core is known to be influenced by the nearby I16293-2422 protostar and the kinetic temperature is significantly higher---increasing from 10~K to 15~K from center to edge \citep{crapsi2007, pagani2013, bacmann2016}---than in the cold cores sampled by \cite{redaelli2018}. Similarly, the temperature in Barnard 1b and OMC-FIR4 is  higher than in the cold cores of their sample \citep{daniel2013, crimier2009}. A trend is thus apparent whereby the isotopic ratio in \nnhp\ decreases as the temperature increases, reaching the bulk ratio of 330. We note that the value of \rx{N2H+} measured toward the three warm} sources is compatible with a value of $\epsilon$ close to 1, which might indicate a temperature dependence of $\epsilon$, from $\approx 2$ below $\sim 10$K toward 1 above $\sim 15$~K. Alternatively, it could be that, in these sources, the destruction of \ce{N2H+} is not dominated by DR with electrons but by other reactions, such as with CO.

Finally, from \rfig{obs} and \rfig{n2hp}, we see that, when $\epsilon>1$, both column density ratios of \rx{N2H+} increase with density, whereas for other species, the ratios decrease from their steady-state value. This could produce some source-to-source variation of the isotopic ratios of different species.

\subsubsection{Ammonia}
\label{sec:govers}
%\revised{13}

Ammonia is, after N and \ce{N2}, an important reservoir of nitrogen in pre-stellar cores and also in cometary ices. Establishing the value of \rrr\ in ammonia at early stages of star formation can therefore bring valuable clues as to the origin of the isotopic ratios measured in comets. The fundamental rotational transition of ammonia cannot be observed in cold gas from the ground. However, direct measurements of \rrr\ in the closely related deuterated variant \ce{NH2D} have been made by \cite{gerin2009b}, using the $1_{1,1}-1_{0,1}$ rotational transition of ortho-\ce{NH2D} at 86~GHz. Despite its very weak intensity, the main hyperfine component of the manifold of \ce{^{15}NH2D} was detected at the 3$\sigma$ level on the integrated intensity toward four sources. The values of \rrr\ extend from 360 up to 810. Despite the large error bars, the depletion in \fifn\ is manifest, as can be seen in \rfig{obs} (see also \rtab{review}). We note that the excitation temperature of \ce{^{15}NH2D} from the LTE analysis of \cite{gerin2009b} is likely to be overestimated \citep{daniel2013}, and so would be the isotopic ratios, by up to a factor of two, bringing down their values toward the bulk isotopic ratio \rtot=330. Assuming a 15\% decrease in the excitation temperature of the \fifn\ variant, relative to the main isotopolog, as suggested by \cite{daniel2013}, we find \rrr=193, 260, and 590, and $>350$, for NGC1333-\ce{DCO+}, L134N(S), L1689N, and L1544, respectively. {Given the large error bars (typically 30\% downward and 30\% to 100\% upward; see Gerin et al.), the ratios in the first two sources would then be marginally compatible with our prediction of no fractionation, as shown in \rfig{obs}.} However, the ratio measured toward L1689N and the lower limit in L1544 indicate depletion of \fifn\, which is not compatible with our results. 

In the above context, we note that the charge transfer reaction between He$^+$ and N$_2$ is another process with a known isotopic effect. This reaction has two output channels,
\begin{align}
	\cee{
		&N2 + He+ -> N2+ + He \\
		&N2 + He+ -> N + N+ + He ,
	}
\end{align}
and has attracted much interest because the energy needed to effect the transition $\rm N_2(X^1\Sigma_g^+) \to N_2^+(C^2\Sigma_u^+)$ nearly equals the helium ionization potential \citep[see][and references therein]{hrodmarsson2019}. In the UGAN network, the total rate constant is $1.2\times 10^{-9}$~cm$^3$s$^{-1}$ and the branching ratio between the dissociative and non-dissociative channels is 1.94, for both N$_2$ isotopologs. However, this branching ratio is known to be sensitive to which isotopolog of \ce{N2} is involved \citep{govers1975}. Room temperature measurements suggest that the dissociative:non-dissociative branching ratios could be 1.5 for $^{14}$N$^{14}$N and 0.69 for $^{14}$N$^{15}$N, with a total rate coefficient of 1.3\tdix{-9}\cccs (Govers, private communication). Once again, theoretical or experimental investigations are required in order to establish the values of these branching ratios at low temperatures.

%We have thus adopted revised BRs suggested by room temperature measurements (Govers et al. 1975 and Govers, private communication): 1.5 for $^{14}$N$^{14}$N and 0.69 for $^{14}$N$^{15}$N. 

We have tested the impact of these branching ratios in our fiducial model (see \rsec{setup} for its parameters). Our results (see \rapp{govers}) indicate that the largest effect is on ammonia and \ce{NH2D}, which see their steady-state isotopic ratios increase to $\approx 450$ and then decrease during the collapse. Thus, it appears that an explanation for a 30\% depletion of ammonia in \fifn\ can be found in the difference in the branching ratios for the reactions of $^{14}$N$^{14}$N and $^{14}$N$^{15}$N with \ce{He+}. We note that, with the new branching ratio, \rrr\ increases also for HCN and CN, while that of \ce{N2H+} decreases, which conflicts with the observations. As to what could lead to the branching ratios varying among sources: Temperature differences might have been considered, as for the factor $\epsilon$ in the case of \ce{N2H+}; but, in the present context, L1544 and L1689N are respectively cold (6~K in the innermost regions) and relatively warm (10-15~K) cores.

\section{Concluding remarks}
\label{sec:conclusions}

This study of the fractionation of nitrogen-containing species is the first to consider the effects of the collapse on the \rrr\ abundance ratios. It was found that observations of CO tightly constrain the characteristic timescale of the collapse to be on the order of the free-fall timescale. Delaying further the collapse would imply more efficient desorption processes than those considered in our study \citep{vasyunin2013}, as our model predicts that CO would be one order of magnitude smaller than is derived from the observations if the timescale was only a factor of two larger than in pure free-fall.

During the quasi-static evolution of the pre-stellar core, the degrees of fractionation of nitrogen-bearing species are modest, as found in earlier studies, and more so when the fractionation reactions and rate coefficients of \citetalias{roueff2015}, rather than those of \citetalias{terzieva2000}, are adopted. We have shown that the differential rates of photodissociation of N$_2$ and $^{15}$NN by the cosmic-ray induced ultraviolet radiation field are crucial to the degree of fractionation of molecular nitrogen, particularly under conditions of chemical equilibrium.

Regarding nitrogen fractionation, the main new result of this study is that, during the gravitational collapse of a pre-stellar core, differential enrichment of the gas in $^{15}$N-containing species occurs owing to their higher mass and hence lower rate of thermal collisions with and adsorption to grains. The increase in the density enhances the effect of depletion-driven fractionation, which can be up to 30\% for species such as atomic nitrogen, CN, and NO. Gas-phase fractionation reactions are relatively insignificant during gravitational collapse, as their associated timescales exceed the free-fall time. For CN, the predicted ratio is in good agreement with the revised observational value toward L1498. We also note that, similar to HCN, the CN/\thcn\ ratio is $\approx 40$, which is substantially lower than the value of \twc/\thc=70 for carbon in the local ISM; this underlines the danger of using the double-isotope-ratio method, discussed in \rapp{review}.

We have explored the influence of different rates for the dissociative recombination of \ce{N2H+} and its \fifn-variants, showing that the observed \textit{depletion} of N$_{2}$H$^+$ in $^{15}$N could be explained if N$^{15}$NH$^+$ and $^{15}$NNH$^+$ dissociatively recombined more rapidly with electrons than $^{14}$N$_{2}$H$^+$. Furthermore, our results suggest that the enhancement of the dissociative recombination rate of the \fifn-variants increases as the temperature decreases. However, high-precision and -accuracy theoretical calculations are required to put this possibility on a firmer footing.

It is probably naive to expect to be able to reproduce the nitrogen isotopic ratios observed in different sources by means of a single model. In fact, there is evidence suggesting that variations of environmental parameters need to be taken into account. Figure~\ref{fig:review} shows that, for \ce{N2H+}, most measurement in pre-stellar cores yield ratios that are $\sim$2--3 times larger than the {bulk} value \rtot=330 for nitrogen; only one measurement is consistent with 330. Further evidence is provided by the interferometric study of \cite{colzi2019}, showing spatial variation of the abundance ratio in a single protostar. In the context of low-mass pre-stellar cores, source-to-source variations---even when located in the same large-scale environment---were also proposed by \cite{magalhaes2018a} to explain the HCN/HC\fifn\ ratio. \cite{hilyblant2017} have noted the large scatter of the directly measured CN/C\fifn\ ratio along various lines of sight through the diffuse ISM \citep{ritchey2015}, with the values being mutually inconsistent by more than 2$\sigma$. Thus, data show that several species---CN, HCN, \ce{N2H+}---exhibit variations that suggest some sensitivity of the isotopic ratio to local physical conditions. In fact, that the environment could induce variations of the \rrr\ abundance ratios is not surprising, since we know that mass fractionation is a temperature-dependent process, and differential photodissociation depends on the cosmic-ray ionization rate and dust size distribution. Therefore, one might contemplate using nitrogen isotopic ratios to constrain the physical parameters, once our understanding of the interstellar nitrogen chemistry and fractionation processes is adequate for this purpose.

\section{Acknowledgments}
We thank the anonymous referee for useful comments and suggestions which helped improving the clarity of this article.
We also thank Laurent Pagani for providing us with the density profile in L183, and Tomas Govers for recommending us the branching ratio between the dissociative and non-dissociative channels in \fifn\foun. This work was supported by the LABEX OSUG@2020 and by the Programme National Chimie Interstellaire (PCMI), and made use of the GRICAD infrastructure, which is supported by Grenoble research communities. DRF acknowledges support from STFC (ST/L00075X/1), including provision of local computing resources.

\bibliographystyle{mnras}
\bibliography{general,chemistry,allbooks,GRENOBLE}

\appendix

\section{Observations}
\label{app:obs}
\subsection{Review of existing measurements}
\label{app:review}

%\begin{figure*}
%	\centering
%	\includegraphics[width=\hsize]{Nisotopic-cores}
%	\caption{Overview of measurements of the nitrogen isotopic ratio in low- to intermediate-mass cores and Class0 (see Table~\ref{tab:review} for details). The horizontal line indicates the value of the isotopic ratio of elemental nitrogen adopted in this study (\rrr=330). Black dots indicate indirect measurements based on the double-isotope method. Colored points refer to direct measurements, with N$_2$H$^{+}$:N$^{15}$NH$^{+}$ in blue. The N$_2$H$^{+}$:$^{15}$NNH$^{+}$ ratios are indicated  by an asterisk.}
%	\label{fig:review}
%\end{figure*}

\def\val#1#2#3{\ensuremath{#1^{+#2}_{-#3}}}
\begin{table*}
	\centering
  \caption{Isotopic ratios of nitrogen-bearing species measured toward low-mass star-forming regions. The ratios have been separated into direct and indirect measurements, with the latter using the double-isotope-ratio method (see also \rfig{review}). Also given are the references used for the density (see \rfig{obs}) when different from that for the isotopic ratio.}
%  	 Unless specified, the N$_2$H$^{+}$ ratio is N$_2$H$^{+}$:N$^{15}$NH$^{+}$.}
  \label{tab:review}
  \begin{tabular}{llllc}
    \toprule
    Species & Value & Source & Reference & Notes\\
    \midrule
		\mc{2}{Indirect measurements}\smallskip\\	
    %&&&&{Indirect measurements}\smallskip\\
    HCN        & 200$\pm$ 40 & IRAS16293A& \cite{wampfler2014} & (1)\\
    HCN        & 315$\pm$ 45 & OMC-3     & \cite{wampfler2014} & (2)\\
    HCN        & 195$\pm$ 55 & L183      & \cite{hilyblant2013a}\\
    HCN        & 250$\pm$110 & L1544     & \cite{hilyblant2013a}\\
    HCN        & 150$\pm$ 50 & L1521E    & \cite{ikeda2002}\\
    HCN        & 250$\pm$ 80 & L1527     & \cite{yoshida2019}\\
    HNC        & 300$\pm$100 & L1527     & \cite{yoshida2019}\\
    \ce{HC3N}  & 257$\pm$ 54 & TMC1(CP)  & \cite{taniguchi2017c}\\
    \ce{HC5N}  & 344$\pm$ 80 & TMC1(CP)  & \cite{taniguchi2017c}\\
    \ce{HC5N}  & 338$\pm$ 12 & L1527     & \cite{araki2016} & (1) \\
    CN         & 510$\pm$ 70 & L1544     & \cite{hilyblant2013a}\\
    CN         & 476$\pm$ 70 & L1498     & \cite{hilyblant2013a}\\
    \midrule
    \mc{2}{Direct measurements} \smallskip\\
    \ce{N2H+}/\nqnhp & \val {920}{300}{200} & L1544        & \cite{bizzocchi2013,redaelli2018} & (3)\\
    \ce{N2H+}/\nqnhp & \val{670}{150}{230}  & L183         & \cite{redaelli2018} & (3)\\
    \ce{N2H+}/\nqnhp & 730$\pm$250          & L429         & \cite{redaelli2018} & (3)\\
    \ce{N2H+}/\nqnhp & \val{580}{140}{110}  & L694-2       & \cite{redaelli2018} & (3)\\
    \ce{N2H+}/\nqnhp & \val{330}{170}{100}  & I16293E      & \cite{daniel2016} & (3)\\
    \ce{N2H+}/\nqnhp & 400$^{+100}_{-65}$   & Barnard 1b   & \cite{daniel2013} & (3)\\
    \ce{N2H+}/\nqnhp & 240$\pm$50           & OMC2-FIR4    & \cite{kahane2018} & (2, 6) \\
    \ce{N2H+}/\qnnhp & \val{1000}{260}{220} & L1544        & \cite{bizzocchi2013,redaelli2018} &\\
    \ce{N2H+}/\qnnhp & \val{700}{210}{140}  & L694-2       & \cite{redaelli2018}, \cite{crapsi2005} & (3)\\
    \ce{N2H+}/\qnnhp & 320$\pm$60           & OMC2-FIR4    & \cite{kahane2018} & (2, 6)\\
    \ce{NH3}         & 334$\pm$50           & Barnard 1b   & \cite{lis2010}\\
    \ce{NH3}         & \val{300}{55}{40}    & Barnard 1b   & \cite{daniel2013} & (3)\\% & \phb{revised}\\
    \ce{NH2D}        & \val{280}{120}{100}  & Barnard 1b   & \cite{daniel2016} & (4)\\% \phb{revised}\\
    \ce{NH2D}        & \val{530}{570}{180}  & L134N(S)     & \cite{gerin2009b}& (5)\\% \phb{revised}\\
    \ce{NH2D} & \val{360}{160}{110} & NGC133-DCO+ & \cite{gerin2009b} & (5)\\
	\ce{NH2D} & \val{810}{600}{250} & L1689N & \cite{gerin2009b} & (5)\\
    HCN              & 338$\pm$28           & L1498        & \cite{magalhaes2018a} & (3)\\
    \ce{HC3N}        & 400$\pm$40           & L1544        & \cite{hilyblant2018b} & (3)\\
    \ce{HC3N}        & 270$\pm$57           & TMC1(CP)     & \cite{taniguchi2017c}\\
    \ce{HC3N}        & 275$\pm$65           & OMC2-FIR4    & \cite{kahane2018} & (2, 6)\\
    \ce{HC5N}        & 323$\pm$ 80          & TMC1(CP)     & \cite{taniguchi2017c}\\
    CN               & \val{240}{135}{65}   & Barnard 1b   & \cite{daniel2013} & (3)\\% & \phb{new}\\
    CN               & 270$\pm$60           & OMC2-FIR4    & \cite{kahane2018} & (2, 6)\\
    CN               & 230$\pm$ 80          & L1527        & \cite{yoshida2019}& (1,7)\\
    CN               & 310$\pm$ 50          & L1498        & This work (see \rapp{cn}) & (3) \\
    \bottomrule
  \end{tabular}
  \tabnotes (1)  Low-mass Class 0. (2)  Intermediate-mass Class 0. (3) Non-LTE calculation. (4) Replaces \cite{gerin2009b} and \cite{daniel2013}. (5) The LTE assumption introduces some bias on the ratio, essentially from the \fifn\ isotopolog column density which may be overestimated \citep[see ][ and \rsec{govers}]{daniel2016}. (6) {An average density $\aver{\nh}=\int_{r_1}^{r_2} 4\pi r^2 n(r) dr/\int_{r_1}^{r_2} 4\pi r^2 dr = 6.8\tdix{5}\ccc$ was computed based on the power-law from \cite{crimier2009}, with $r_1=100$~au and $r_2=\dix{4}$ au or 25\arcsec\ at 400\,\pc. (7) Same as (6) but with the power-law density profile from \citep{shirley2002}, with $r_1=1000$~au and $r_2=2100$ au or 15\arcsec\ at 140~pc. Assuming an abundance of 10\% for Helium, the average density is $\aver{\nh}=6.5\tdix{5}$\ccc.}
\end{table*}

%	195$\pm$55 & HCN        & L183      & \cite{hilyblant2013a}\\
%	200$\pm$40 & HCN        & IRAS16293A& \cite{wampfler2014} & Low-mass Class 0\\
%	315$\pm$45 & HCN        & OMC-3     & \cite{wampfler2014} & Intermediate-mass Class 0\\
%	250$\pm$110 & HCN        & L1544     & \cite{hilyblant2013a}\\
%	257$\pm$54 & HC$_3$N  & TMC1(CP)  & \cite{taniguchi2017c}\\
%	344$\pm$80 & HC$_5$N  & TMC1(CP)  & \cite{taniguchi2017c}\\
%	150$\pm$50 & HCN        & L1521E    & \cite{ikeda2002}\\
%	338$\pm$12 & HC$_5$N  & L1527     & \cite{araki2016}\\
%	510$\pm$70 & CN         & L1544     & \cite{hilyblant2013a}\\
%	476$\pm$70 & CN         & L1498     & \cite{hilyblant2013a}\\

We present, in Table~\ref{tab:review}, an overview of measurements of the nitrogen isotopic ratio in pre-stellar cores, derived from observations of different molecular species. Direct and indirect measurements are shown separately. We recall that indirect measurements are based on the so-called double-isotope method, used to circumvent the problem of allowing for the optical depth in the line of the main isotopolog. For instance, one could use a \thc\ substitute to the \twc\ species
\[
\rm H\twc N/H\twc\fifn = (H\twc N/H\thc N) \times H\thc N / H\twc\fifn
\]
or a deuterated variant
\[
\rm H\twc N/H\twc \fifn = (H\twc N/D\twc N) \times D\twc N / H\twc\fifn
\]
but one has to assume a value for the term in brackets, which is not measured. The usual assumption consists of adopting the isotopic ratio of the element, unless there is clear evidence for systematic deviation. Thus, the (H\twc N/H\thc N) term is usually taken equal to \twc/\thc=70 in the local ISM. However, such an approximation is not always justified and may lead to error. 

Direct measurements can be made by several means. \cite{gerin2009b} proposed using the deuterated form of \fifn-ammonia to measure the nitrogen isotopic ratio directly through the ratio \ce{NH2D}/\ce{^{15}NH2D}. A more general way of obtaining direct ratios was put forward by \cite{daniel2013}, based on radiative transfer calculations with the \alico\ code, which takes into account the contribution of overlap of the hyperfine transitions in the excitation of the hyperfine levels. This approach was further developed by \cite{magalhaes2018a}, who combined \alico\ with a Markov-chain Monte Carlo (MCMC) exploration of the parameter space. This method enabled the first direct measurement of \rrr\ for HCN in a pre-stellar core; the result was found to differ from estimates based on the double-isotopic-ratio method. Direct measurements in pre-stellar cores have been made by various groups, as reported in Table~\ref{tab:review}.

An illustration of the importance of making direct rather than indirect measurements of the degree of fractionation is shown in Fig.~\ref{fig:review} (see also Table~\ref{tab:review}). {The indirect ratios span a broad range of values, from 150 to 510, with an average 283$\pm$93, while direct measurements for the same species (nitriles) are less scattered, with an average 295$\pm$50. Nevertheless, the $1/\sigma^2$-weighted mean values of indirect (\aver{\rrr}=314$\pm$10) and direct measurements (\aver{\rrr}=329$\pm$14) are consistent within $1\sigma$\footnote{The uncertainty of the weighted average is estimated as $(\sum_k \sigma_k^{-2})^{-1/2}$ and is quoted at the 1$\sigma$ level.}. We note that the uncertainty of the \cite{araki2016} indirect measurement is most likely underestimated. Yet, the weighted averages are} in remarkable agreement with the CN/C$^{15}$N ratio measured directly in the disk orbiting TW~Hya, and also with the prediction of the galactic chemical evolution model of \cite{romano2017}, lending support to the value of 330 adopted in this study for the present-day $^{14}$N:$^{15}$N ratio \citep{hilyblant2017, hilyblant2019}. We also note that the direct and indirect averages {are} consistent within $1\sigma$.

From Fig.~\ref{fig:review}, one also notices that the \nnhp\ ratios, all obtained with the direct method in a sample of starless cores \citep{bizzocchi2013, redaelli2018}, are significantly higher than in other species. Yet, toward the relatively warmer OMC2-FIR4 and I16293E sources, the observed ratios agree well with 330. Indeed, the weighted average of all \nnhp\ ratios is 350$\pm34$ (we note that unweighted average value is 610$\pm$250). Furthermore, when excluding \nnhp, the weighted average of direct measurements becomes 325$\pm15$.

%An exception is the value of 365$\pm$135 in the pre-stellar core I16293E \citep{daniel2016} that is interacting with the nearby Class 0 source I16293-2422, which has a central temperature larger than other starless cores \citep{crapsi2007, pagani2013}.

\subsection{Re-evaluation of the CN/C\fifn\ ratio toward L1498}
\label{app:cn}

\begin{figure}
	\centering
	\includegraphics[width=\hsize]{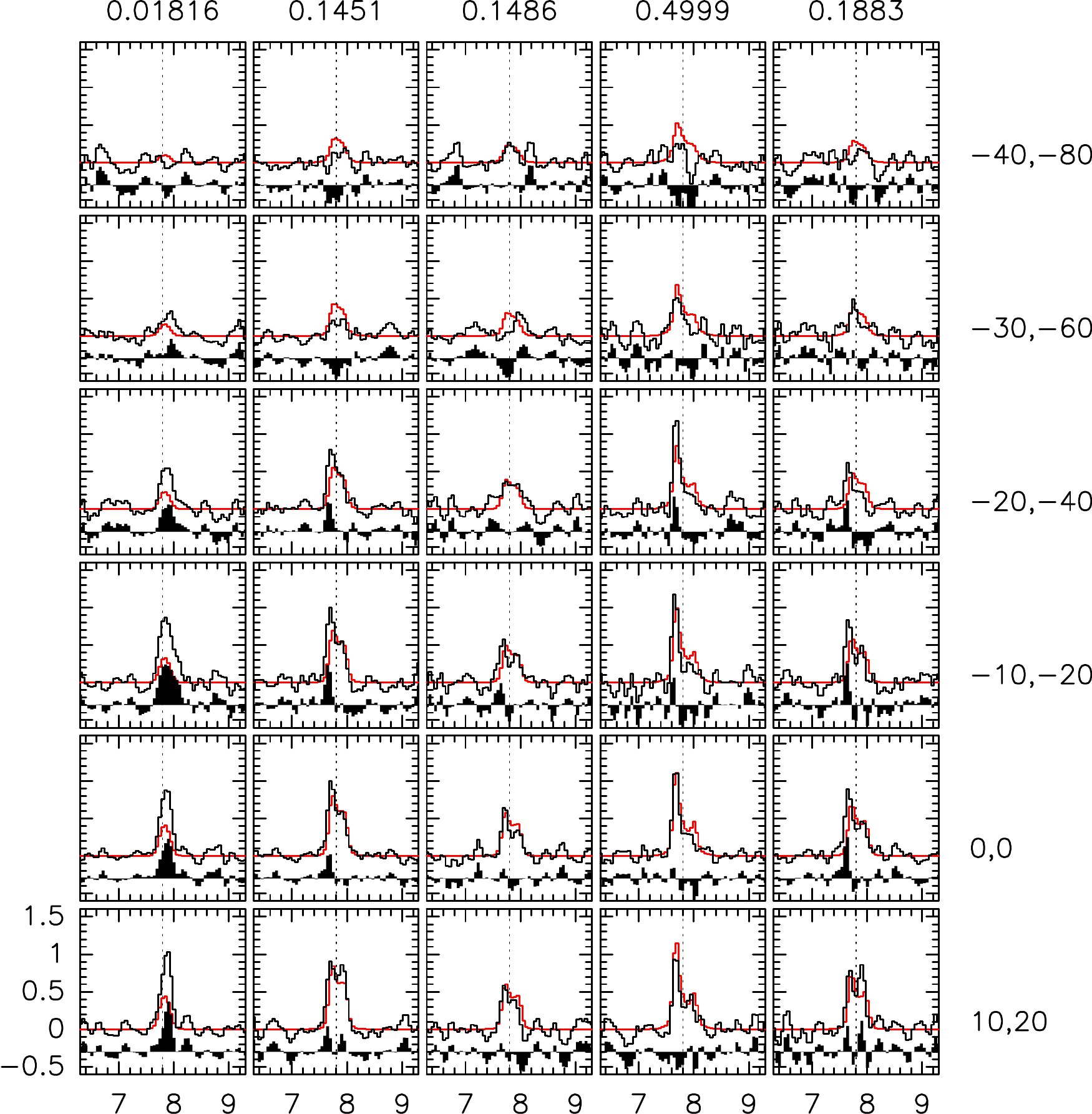}\bigskip\\
	\includegraphics[width=0.8\hsize]{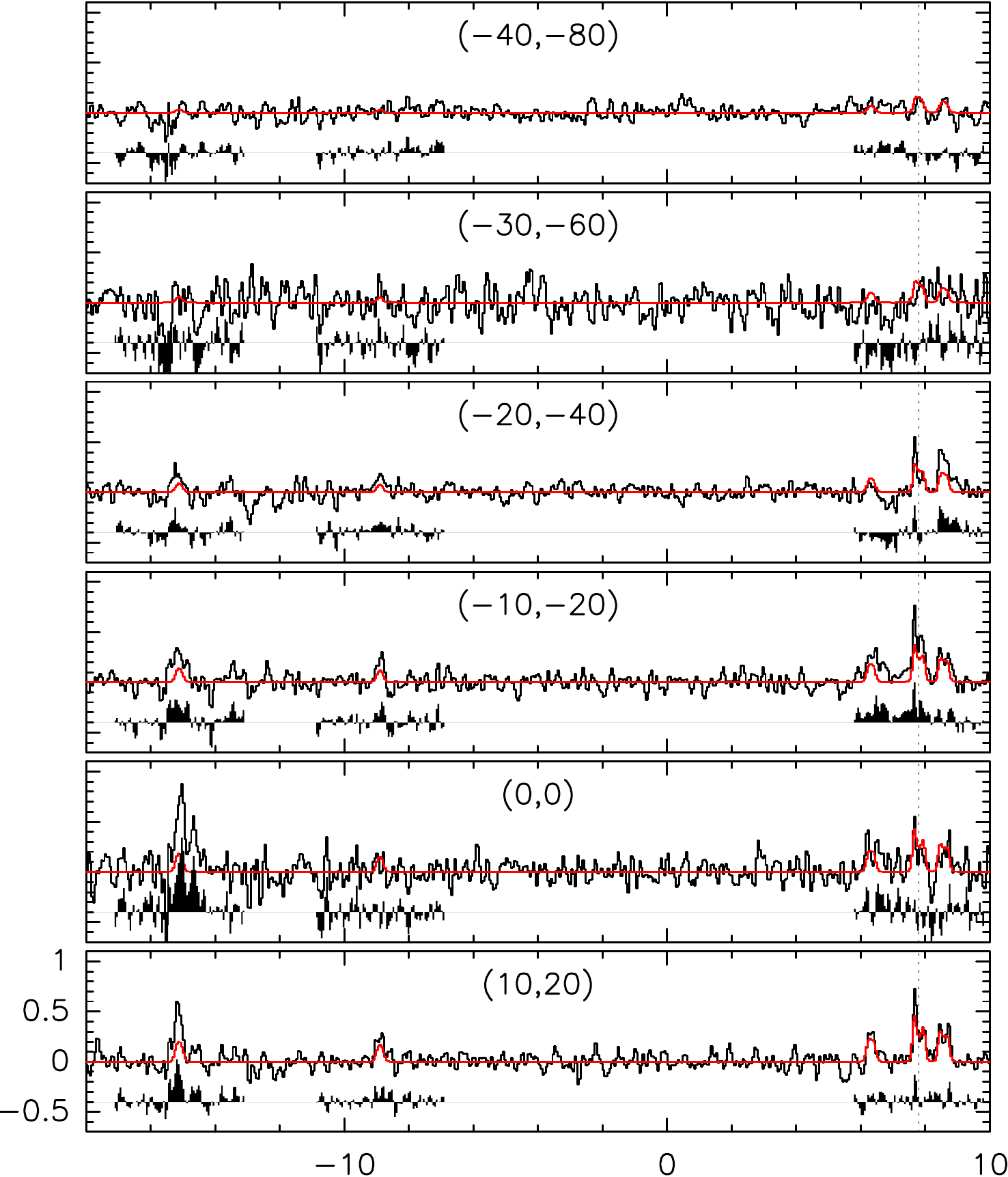}
	\caption{The observed hyperfine manifolds of CN(1-0) at 113.5~GHz (top) and of CN(2-1) at  (bottom) toward L1498 as measured with the IRAM-30m telescope \citep{hilyblant2010n}. For CN(1-0), the five hyperfine lines are shown separately, with their relative intensity indicated at the top. The best-fit model is shown in red, and the residuals are shown below each spectrum. The offsets are indicated on the right of each panel. The x-axis is the LSR velocity, and the intensity is on a main-beam temperature scale.}
	\label{fig:12cn}
\end{figure}

As was emphasized above, direct measurements of isotopic ratios are essential because of the unquantifiable uncertainties introduced when using the double-isotope method \citep{gerin2009b, hilyblant2017}. In the L1498 pre-stellar core, the CN/C\fifn\ ratio was determined using the latter method \citep{hilyblant2013b}, in order to circumvent the optical-depth of the main isotopolog. The derived ratio was \rrr=476$\pm$70, where the error was only statistical and did not include the uncertainty intrinsic to the double-isotope method. To obtain a direct measurement, we used the source model of \cite{magalhaes2018a} and followed the same methodology and assumptions. The minimization, performed using Markov chains, was done simultaneously on the CN, \thcn, and C\fifn, spectra obtained along cuts across the core (see \rfigs{12cn}{13cn_c15n}). The inner plateau of constant density has a radius of (77$_{-10}^{+5}$)\arcsec, which is larger than the value of 47$\pm6$\arcsec\ found for HCN; this appears to indicate that CN is more extended than HCN, but a comprehensive study of L1498, similar to that performed by \cite{daniel2013} for Barnard1b, is beyond the scope of the present study. The resulting relative abundances are log$_{10}$(CN/\hh)=$-8.51_{-0.07}^{0.08}$,  log$_{10}$(CN/\thcn)=$1.59_{-0.06}^{0.08}$, and  log$_{10}$(C\fifn/CN)=$2.49_{-0.06}^{0.07}$ (see \rfig{corner}), corresponding to CN/\hh=3.2$\pm$0.5\tdix{-9}, CN/C\fifn=310$_{-40}^{53}$, and CN/\thcn=39$_{-4}^{+8}$.

%   Burning          Initial values            Plateau size         log10(CN/H2)           log10(CN/13CN)        log10(CN/C15N)
%...........................
%50     0.1    -8.6     1.7     2.5      0.16 -0.01  0.01     -8.52 -0.07  0.08      1.59 -0.07  0.08      2.48 -0.07  0.07
%100     0.1    -8.6     1.7     2.5      0.16 -0.01  0.01     -8.51 -0.07  0.08      1.59 -0.06  0.08      2.49 -0.06  0.07
%140     0.1    -8.6     1.7     2.5      0.16 -0.02  0.01     -8.50 -0.07  0.08      1.60 -0.06  0.07      2.49 -0.06  0.07
%
%CN/C15N = 310  270 363
%CN/13CN = 39    35  47
%CN/H2   = 3.2  2.7 3.8 x 1e-9
%plateau = 0.16 0.14 0.17 or 77" 67" 82"

\begin{table}
	\centering
	\caption{\label{tab:c15n}Spectroscopic properties of the transitions of the CN(1-0) and (2-1) hyperfine manifolds considered in \rapp{cn}.}
	\begin{tabular}{rrrrrr}
		\toprule
		$\delta v$ & R.I. & Rest Freq. & \aul & $g_u$ \\
		\kms & & MHz & \pers \\
		\midrule
		CN(1-0)\\
		30.439  & 1.883E-01 & 113488.120200  & 6.736E-06  &  4 \\
		22.911   &4.999E-01 & 113490.970200  & 1.192E-05  &  6 \\
		0.000   &1.486E-01 & 113499.644300   &1.063E-05  &  2   \\
		-24.467  & 1.451E-01&  113508.907400  & 5.190E-06  &  4\\
		-54.906   &1.816E-02&  113520.431500   &1.300E-06 &   2\\
		CN(2-1)\\
		1.473&  4.454E-01& 226874.781300 &  1.143E-04&    8\\
		-15.228&  5.321E-02& 226887.420200 &  2.731E-05&    4\\
		-21.449&  5.291E-02& 226892.128000 &  1.811E-05&    6\\
		\bottomrule
	\end{tabular}
\end{table}

\begin{figure*}
	\centering
	\includegraphics[width=.45\hsize]{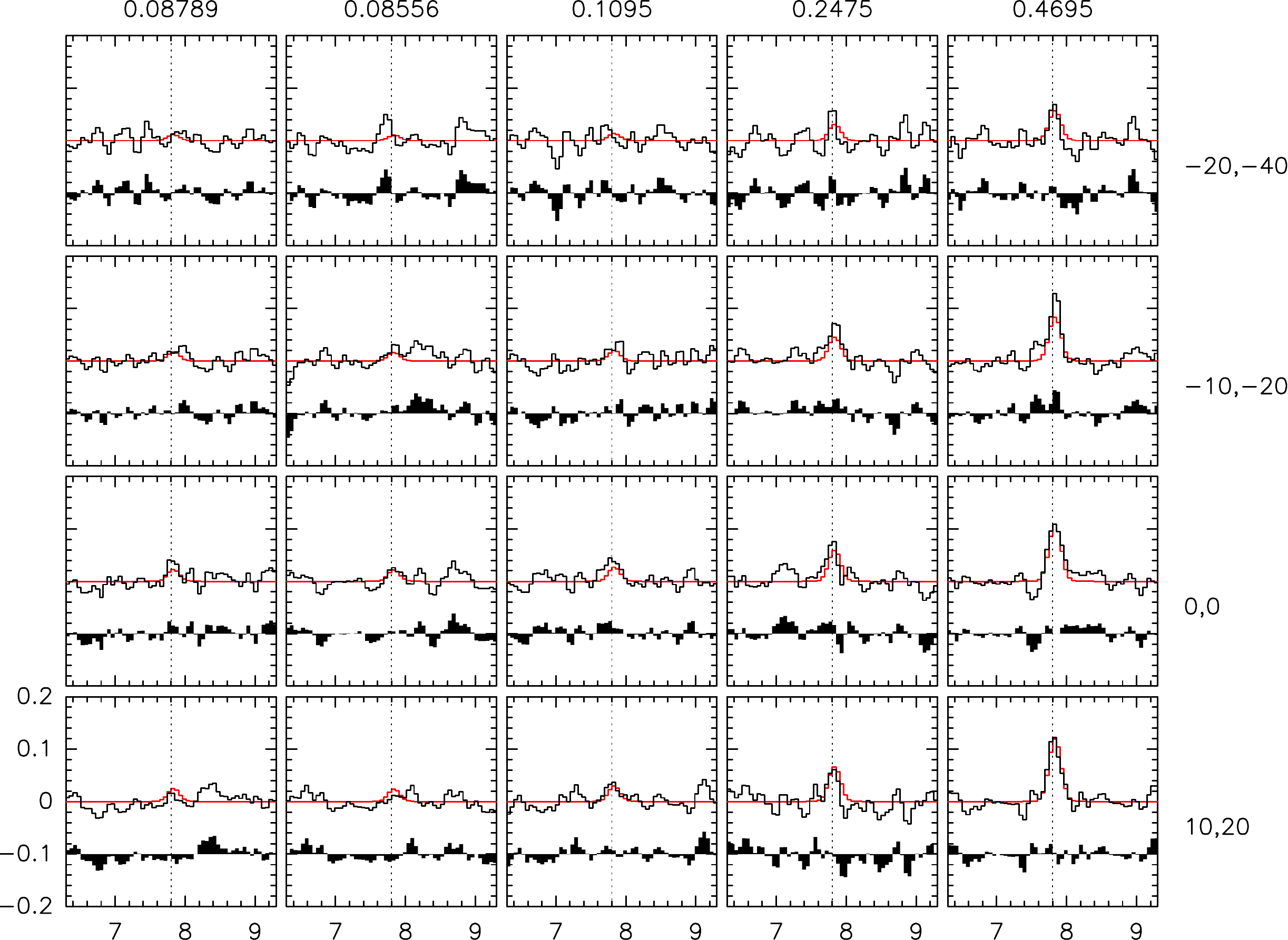}\hfill% bigskip\\
	\includegraphics[width=.45\hsize]{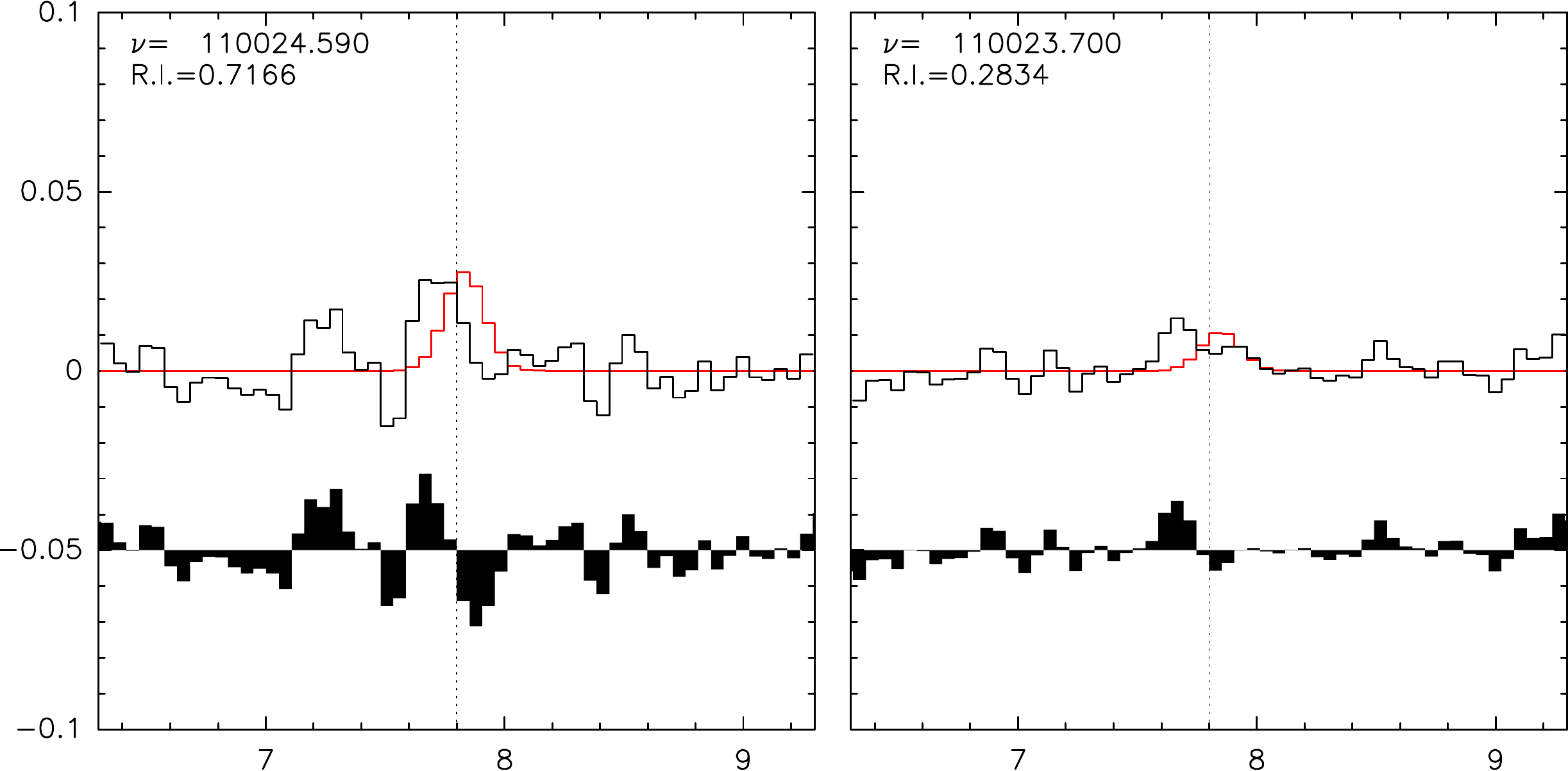}
	\caption{Same as \rfig{12cn} for \thcn(1-0) (left) and C\fifn(1-0) (right).}
	\label{fig:13cn_c15n}
\end{figure*}
\begin{figure}
	\centering
	\includegraphics[width=\hsize]{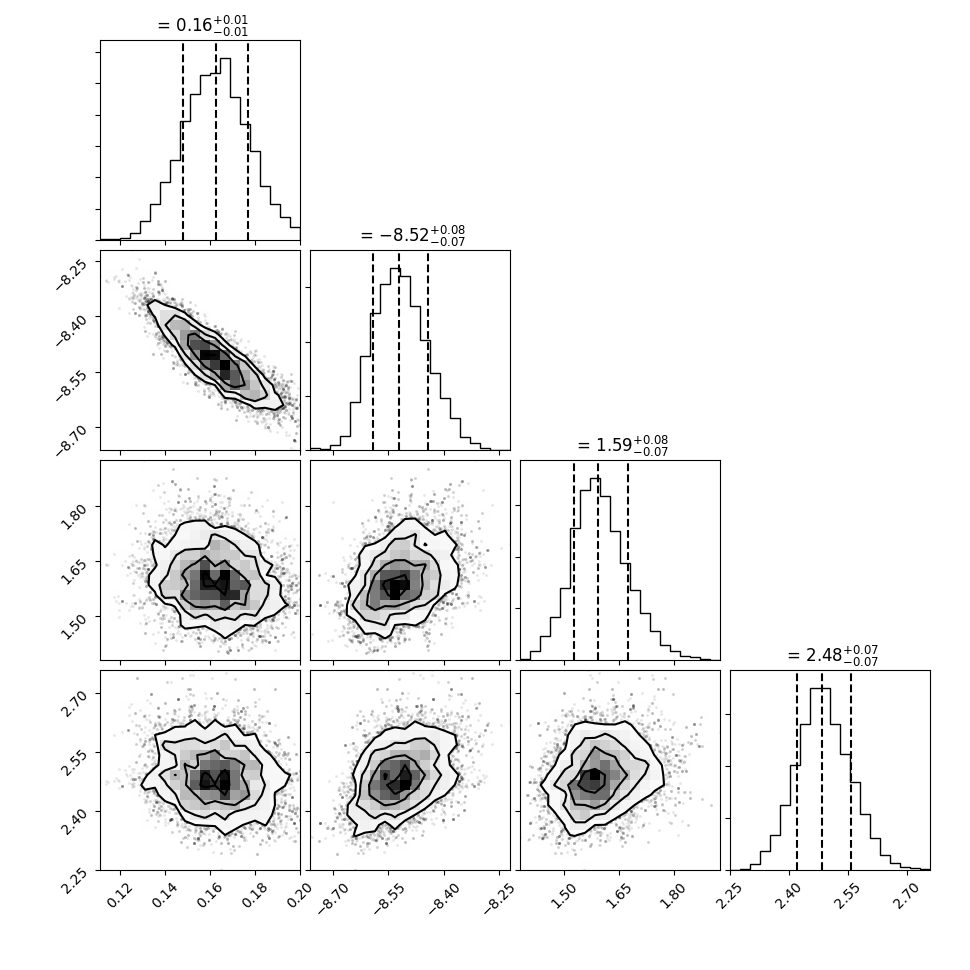}
	\caption{Correlations between the four parameters {(from left to right): the radius (expressed as a fraction of 8\arcmin) of the constant density plateau in the core density profile, log$_{10}$(CN/\hh), log$_{10}$(CN/$^{13}$CN), and log$_{10}$(CN/C$^{15}$N).} The first 50 MCMC steps were not included.}
	\label{fig:corner}
\end{figure}

%\section{Astrochemical details}
\section{Building the $^{15}$N network}
\label{app:network}
{Most reactions can be duplicated in an automatic way, and routine was developed to construct the \fifn- from the \foun-network. But some reactions must be treated individually. We recall that our network is limited to singly substituted species.

Reactions such as
\begin{align}
\cee{
	&N     +    crp  -> N+      + e-\\
	&NH  +    O      -> OH     + N\\
	&NO  +    N      -> N2      + O
	}
\end{align}
(with "crp" standing for cosmic ray particles) are simply duplicated. This is also the case for adsorption and desorption reactions (see \rtab{grain_ads} and \rtab{grain_des}) where equal binding energies are adopted for the \foun\ and \fifn\ species.

In other instances, such as
\begin{equation}\label{key}
\ce{N2 + He+ -> N+ + N + He},
\end{equation}
the corresponding \fifn\ reaction rate coefficient is assumed to be divided equally among the output channels (in this example, each channel is attributed half of the \ce{N2} reaction rate coefficient). Finally, for the following three reactions, which we assume to proceed through proton- and deuteron-hop \citep{hemsworth1974}, the rate coefficients of each of the \fifn-substituted reaction are taken equal to the corresponding \foun\ rate coefficient:
\begin{align}
%%\ce{N2H+ +  NH3  -> NH4+ + N2} \label{equ3}&&\\
%%\ce{N2D+ + NH3 -> NH3D+ + N2}\label{equ4}\\
%%\ce{N2H+ + NO -> HNO+ + N2}\label{equ5}
\cee{
	&N2H+ +  NH3 -> NH4+ + N2 \\
	&N2D+ + NH3 -> NH3D+ + N2\\
	&N2H+ + NO -> HNO+ + N2}
\end{align}

The fractionation reactions of \citetalias{terzieva2000} and \citetalias{roueff2015} are listed in \rtab{frac}.

%assuming identical branching ratios for the corresponding output channels, and has been limited to singly-substituted species. The following reactions were treated manually assuming that they proceed by proton-hop and deuteron-hop:
 %(see \rtab{hop}).

\begin{table}
	\caption{List of adsorption reactions.  The \fifn\ substitutions, which are identical to the \foun\ reactions, are not shown here.}
	\label{tab:grain_ads}
	\footnotesize
	\begin{tabular}{l@{+\,}l@{\qquad \ra\qquad}lllrrr}
  \toprule
C        & GRAIN &     CH4*  &         &                        \\
CH       & GRAIN &     CH4*  &         &                        \\
CH2      & GRAIN &     CH4*  &         &                        \\
CH3      & GRAIN &     CH4*  &         &                        \\
CH4      & GRAIN &     CH4*  &         &                        \\
O        & GRAIN &     H2O*  &         &                        \\
OH       & GRAIN &     H2O*  &         &                        \\
H2O      & GRAIN &     H2O*  &         &                        \\
N        & GRAIN &     NH3*  &         &                        \\
NH       & GRAIN &     NH3*  &         &                        \\
NH2      & GRAIN &     NH3*  &         &                        \\
NH3      & GRAIN &     NH3*  &         &                        \\
S        & GRAIN &     H2S*  &         &                        \\
SH       & GRAIN &     H2S*  &         &                        \\
H2S      & GRAIN &     H2S*  &         &                        \\
CO       & GRAIN &     CH3OH*&         &                        \\
H2CO     & GRAIN &     CH3OH*&         &                        \\
CH3OH    & GRAIN &     CH3OH*&         &                        \\
O2       & GRAIN &     O2*   &         &                        \\
CO2      & GRAIN &     CO2*  &         &                        \\
C2       & GRAIN &     CH4*  &   CH4*  &                        \\
C2H      & GRAIN &     CH4*  &   CH4*  &                        \\
C2H2     & GRAIN &     CH4*  &   CH4*  &                        \\
C3       & GRAIN &     CH4*  &   CH4*  &   CH4*                 \\
C3H      & GRAIN &     CH4*  &   CH4*  &   CH4*                 \\
C3H2     & GRAIN &     CH4*  &   CH4*  &   CH4*                 \\
CN       & GRAIN &     CH4*  &   NH3*  &                        \\
HCN      & GRAIN &     CH4*  &   NH3*  &                        \\
HNC      & GRAIN &     CH4*  &   NH3*  &                        \\
N2       & GRAIN &     N2*   &         &                        \\
NO       & GRAIN &     H2O*  &   NH3*  &                        \\
CS       & GRAIN &     CH4*  &   H2S*  &                        \\
SO       & GRAIN &     H2O*  &   H2S*  &                        \\
SO2      & GRAIN &     H2O*  &   H2O*  &   H2S*                 \\
OCS      & GRAIN &     OCS*  &         &                        \\
Fe       & GRAIN &     Fe*   &         &                        \\
OD       & GRAIN &     HDO*  &         &                        \\
HDO      & GRAIN &     HDO*  &         &                        \\
D2O      & GRAIN &     D2O*  &         &                        \\
ND       & GRAIN &     NH2D* &         &                        \\
NHD      & GRAIN &     NH2D* &         &                        \\
ND2      & GRAIN &     NHD2* &         &                        \\
NH2D     & GRAIN &     NH2D* &         &                        \\
NHD2     & GRAIN &     NHD2* &         &                        \\
ND3      & GRAIN &     ND3*  &         &                        \\
  \bottomrule
\end{tabular}
\tabnote Mantle species are denoted by an asterisk. Grain-surface chemistry in the UGAN network is handled in a simplified way whereby sticking atoms become saturated with hydrogen. This amounts to assuming that the timescale for the formation of saturated molecules in ices is much shorter than the other chemical timescales.

%%% Local Variables:
%%% mode: latex
%%% TeX-master: t
%%% End:

\end{table}

\begin{table*}
	\caption{List of direct cosmic-ray desorption reactions. The \fifn\ substitutions, which are identical to the \foun\ reactions, are not shown here.}
	\label{tab:grain_des}
	\centering
	\footnotesize
	\begin{tabular}{l@{+\,}l@{\qquad \ra\qquad}llcc}
  \toprule
%\multicolumn{5}{l}{Direct cosmic-ray desorption$^\S$}\\
\multicolumn{4}{l}{Reaction} & $\alpha^{(1)}$ & $E_{\rm ads}$\\
%(1) &(2) &(3) &(4) &(5) &(6) &(7)\\
\midrule
CH4*     & CRP  &      CH4    &  GRAIN &                        7.000e+01&   1.120e+03\\
O2*      & CRP  &      O2     &  GRAIN &                        7.000e+01&   1.210e+03\\
H2O*     & CRP  &      H2O    &  GRAIN &                        8.000e-01&   8.550e+02\\
CO*      & CRP  &      CO     &  GRAIN &                        7.000e+01&   8.550e+02\\
CO2*     & CRP  &      CO2    &  GRAIN &                        7.000e+01&   2.690e+03\\
N*       & CRP  &      N      &  GRAIN &                        7.000e+01&   8.000e+02\\
NH3*     & CRP  &      NH3    &  GRAIN &                        2.360e+00&   8.550e+02\\
N2*      & CRP  &      N2     &  GRAIN &                        7.000e+01&   7.900e+02\\
CH3OH*   & CRP  &      CH3OH  &  GRAIN &                        7.000e+01&   4.240e+03\\
H2CO*    & CRP  &      H2CO   &  GRAIN &                        7.000e+01&   1.760e+03\\
HCO2H*   & CRP  &      HCO2H  &  GRAIN &                        7.000e+01&   1.500e+03\\
OCS*     & CRP  &      OCS    &  GRAIN &                        7.000e+01&   3.000e+03\\
H2S*     & CRP  &      H2S    &  GRAIN &                        7.000e+01&   1.800e+03\\
Fe*      & CRP  &      Fe     &  GRAIN &                        7.000e+01&   2.400e+03\\
HDO*     & CRP  &      HDO    &  GRAIN &                        8.000e-01&   8.550e+02\\
D2O*     & CRP  &      D2O    &  GRAIN &                        8.000e-01&   8.550e+02\\
NH2D*    & CRP  &      NH2D   &  GRAIN &                        2.360e+00&   8.550e+02\\
NHD2*    & CRP  &      NHD2   &  GRAIN &                        2.360e+00&   8.550e+02\\
ND3*     & CRP  &      ND3    &  GRAIN &                        2.360e+00&   8.550e+02\\
%\midrule
%\multicolumn{5}{l}{Indirect cosmic-ray desorption}\\
%CH4*     & SECPHO&     GRAIN&    CH4   &                        \\
%O2*      & SECPHO&     GRAIN&    O2    &                        \\
%H2O*     & SECPHO&     GRAIN&    H2O   &                        \\
%CO*      & SECPHO&     GRAIN&    CO    &                        \\
%CO2*     & SECPHO&     GRAIN&    CO2   &                        \\
%N*       & SECPHO&     GRAIN&    N     &                        \\
%NH3*     & SECPHO&     GRAIN&    NH3   &                        \\
%ND3*     & SECPHO&     GRAIN&    ND3   &                        \\
%N2*      & SECPHO&     GRAIN&    N2    &                        \\
%CH3OH*   & SECPHO&     GRAIN&    CH3OH &                        \\
%H2CO*    & SECPHO&     GRAIN&    H2CO  &                        \\
%HCO2H*   & SECPHO&     GRAIN&    HCO2H &                        \\
%OCS*     & SECPHO&     GRAIN&    OCS   &                        \\
%H2S*     & SECPHO&     GRAIN&    H2S   &                        \\
%Fe*      & SECPHO&     GRAIN&    Fe    &                        \\
%HDO*     & SECPHO&     GRAIN&    HDO   &                        \\
%D2O*     & SECPHO&     GRAIN&    D2O   &                        \\
%NH2D*    & SECPHO&     GRAIN&    NH2D  &                        \\
%NHD2*    & SECPHO&     GRAIN&    NHD2  &                        \\
  \bottomrule
\end{tabular}
\tabnotes (1) The rate is computed as in \cite{flower2005} (see their Section 3.3) and \cite{faure2019} (Eq. 7), where the yield (noted $\gamma$ in \citeauthor{flower2005}) is given by $\alpha\zeta_\hh$, and the binding energy is $E_{\rm ads}$. The binding energies of the \foun\ and \fifn\ variants of adsorbed species have equal values. The same desorption reactions, but by secondary photons (photodesorption), are also taken into account in UGAN. (2) CRP stands for cosmic-ray particle. (3) Mantle species are denoted by an asterisk.

%%% Local Variables:
%%% mode: latex
%%% TeX-master: t
%%% End:

\end{table*}

\begin{table*}
	\centering
	\caption{The fractionation reactions of \citetalias{terzieva2000} (top) and \citetalias{roueff2015} (bottom). The rate is computed as in \req{rate}.}
	\label{tab:frac}
	\footnotesize
	\begin{tabular}{lrrrc}
  \toprule
  Reaction& $\alpha$ & $\beta$ & $\gamma$ &  Comment \\
  & \cccs    &  &K\\
  \midrule
%  \cite{terzieva2000}\\
  \ce{^{15}NN + N2H+   -> N2    + N^{15}NH+ }& 1.00E-09 &  0.00E+00 & 0.00E+00 \\
  \ce{N2  + N^{15}NH+  -> ^{15}NN   + N2H+  }& 2.00E-09 &  0.00E+00 & 1.07E+01 \\
  \ce{^{15}NN + N2H+   -> N2    + ^{15}NNH+ }& 1.00E-09 &  0.00E+00 & 0.00E+00 \\
  \ce{N2  + ^{15}NNH+  -> ^{15}NN   + N2H+  }& 2.00E-09 &  0.00E+00 & 2.25E+00 \\
  \ce{^{15}N+ + N2     -> N+    + ^{15}NN   }& 1.00E-09 &  0.00E+00 & 0.00E+00 \\
  \ce{N+  + ^{15}NN    -> ^{15}N+   + N2    }& 5.00E-10 &  0.00E+00 & 2.83E+01 \\
  \ce{^{15}N+ + NO     -> N+    + ^{15}NO   }& 1.00E-09 &  0.00E+00 & 0.00E+00 \\
  \ce{N+  + ^{15}NO    -> ^{15}N+   + NO    }& 1.00E-09 &  0.00E+00 & 2.43E+01 \\
  \ce{^{15}N  + N2H+   -> ^{15}NNH+ + N     }& 1.00E-09 &  0.00E+00 & 0.00E+00 \\
  \ce{N   + ^{15}NNH+  -> ^{15}N    + N2H+  }& 1.00E-09 &  0.00E+00 & 3.61E+01 \\
  \ce{^{15}N  + N2H+   -> N^{15}NH+ + N     }& 1.00E-09 &  0.00E+00 & 0.00E+00 \\
  \ce{N   + N^{15}NH+  -> ^{15}N    + N2H+  }& 1.00E-09 &  0.00E+00 & 2.77E+01 \\
  \ce{^{15}N  + HCNH+  -> N     + HC^{15}NH+}& 1.00E-09 &  0.00E+00 & 0.00E+00 \\
  \ce{N   + HC^{15}NH+ -> ^{15}N    + HCNH+ }& 1.00E-09 &  0.00E+00 & 3.59E+01 \\
  \ce{^{15}N  + H2NC+  -> N     + H2^{15}NC+}& 1.00E-09 &  0.00E+00 & 0.00E+00 & (1) \\
  \ce{N   + H2^{15}NC+ -> ^{15}N    + H2NC+ }& 1.00E-09 &  0.00E+00 & 3.59E+01 & (1) \\
  \midrule
%  \cite{roueff2015}\\
  \ce{^{15}NN + N2H+   -> N2    + N^{15}NH+ }& 2.30E-10 &  0.00E+00 & 0.00E+00 \\
  \ce{N2  + N^{15}NH+  -> ^{15}NN   + N2H+  }& 4.60E-10 &  0.00E+00 & 1.03E+01 \\
  \ce{^{15}NN + N2H+   -> N2    + ^{15}NNH+ }& 2.30E-10 &  0.00E+00 & 0.00E+00 \\
  \ce{N2  + ^{15}NNH+  -> ^{15}NN   + N2H+  }& 4.60E-10 &  0.00E+00 & 2.10E+00 \\
  \ce{^{15}NN + ^{15}NNH+  -> ^{15}NN   + N^{15}NH+ }& 4.60E-10 &  0.00E+00 & 0.00E+00 \\
  \ce{^{15}NN + N^{15}NH+  -> ^{15}NN   + ^{15}NNH+ }& 4.60E-10 &  0.00E+00 & 8.10E+00 \\
  \ce{^{15}N+ + N2     -> N+    + ^{15}NN   }& 4.80E-10 &  0.00E+00 & 0.00E+00 \\
  \ce{N+  + ^{15}NN    -> ^{15}N+   + N2    }& 2.40E-10 &  0.00E+00 & 2.83E+01 \\
  \ce{^{15}N  + CN     -> N     + C^{15}N   }& 0.00E-10 &  1.667E-01 & 0.00E+00 & (2)\\
  \ce{N   + C^{15}N    -> ^{15}N    + CN    }& 0.00E-10 &  1.667E-01 & 2.29E+01 & (2)\\
  \ce{^{15}N  + C2N+   -> N     + C2^{15}N+ }& 3.80E-12 & -1.00E+00 & 0.00E+00 \\
  \ce{N   + C2^{15}N+  -> ^{15}N    + C2N+  }& 3.80E-12 & -1.00E+00 & 3.81E+01 \\
  \bottomrule
\end{tabular}
\tabnotes (1) Added for consistency with the UGAN network. (2) Our fiducial network does not take into account the possibility that CN + N could lead to isotopic exchange \citep[see also][]{wirstrom2018}.

%%% Local Variables:
%%% mode: latex
%%% TeX-master: t
%%% End:

\end{table*}

%\begin{table*}
%	\caption{The fractionation reactions of \citetalias{terzieva2000} and \citetalias{roueff2015}. The rate is computed as in \req{rate}.}
%	\label{tab:hop}
%	\footnotesize
%	\input{protonhop_table}
%\end{table*}
}

\section{Depletion-driven fractionation}
\label{app:depletion}
\begin{figure*}
	\centering
	\includegraphics[width=\hsize]{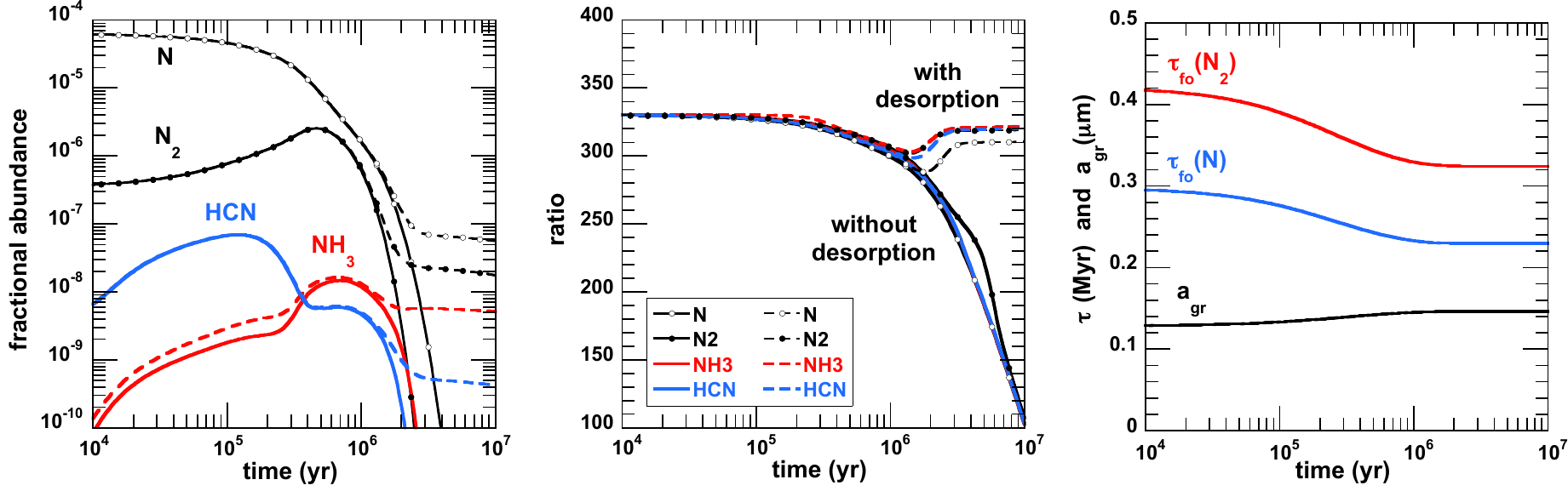}
	\caption{\textit{Left panel:}Temporal evolution of the abundances of N, \ce{N2}, HCN, and \ce{NH3} in a constant density model with our fiducial parameters {(see \rsec{setup})} and using the  \citetalias{roueff2015} rates. \textit{Middle panel:} Same for the isotopic ratios. The full curves are for models with adsorption only, while dashed curves are for models with both adsorption and desorption. \textit{Right panel:} Temporal evolution of the freeze-out timescales $\tau_{\rm fo}$ of N and \ce{N2} and of the dust grain radius \agr.}
	\label{fig:depletion}
\end{figure*}

The freeze-out timescale for a species of mass $m=\mamu\mh$\footnote{We neglect the difference between \mh\ and the atomic mass unit.} is
\begin{equation}
\taufo = (n_d \pi \agr^2 \vther S)^{-1},
\label{eq:taufo_def}
\end{equation}
where $n_d$ is the dust density [\ccc] and $S$ is the sticking coefficient, which is taken equal to 1 for all species in our isothermal models at 10~K. For a Maxwell-Boltzmann velocity distribution, the average thermal speed, $v_{\rm th}$, is given by $(8kT/\pi m)^{1/2}$.

In our model, grains are decomposed into a core, made of refractory material and of constant composition, and mantles which result from the competition between adsorption and desorption processes. The grain size, \agr, is thus the radius of the core, \acore, plus mantles. Noting $X_m$ ($X_c$) the fractional abundance of a mantle (core) species, and $A_m$ ($A_c$) its atomic mass number, the grain radius (core and mantles) is given by
\begin{equation}
\label{eq:agr}
\agr=\acore\lp 1+\frac{\sum_m X_m A_m}{\sum_c X_c A_c}\frac{\rho_c}{\rho_m}\rp^{1/3},
\end{equation}
where the core and mantle mass density are $\rho_c=2$\gccc\ and $\rho_m=1$\gccc, respectively. With the core and mantles composition listed in \rtab{abinit}, the initial grain radius is $\agr=1.28 \acore$ or 0.128\micr. We note that the refractory:gas mass ratio (usually called dust:gas mass ratio) is
\begin{equation}
\label{eq:qc}
Q = \sum_c X_c A_c/\mu=0.0061,
\end{equation}
while the grain:gas mass ratio is
\begin{equation}
\label{eq:qm}
Q_m = \lp\sum_c X_c A_c + \sum_m X_m A_m \rp/\mu=0.0094,
\end{equation}
where $\mu=1.4$ is the mean molecular weight per hydrogen nucleus. The fractional dust abundance is then:
\begin{equation}
\label{eq:xd}
X_d = \frac{n_d}{\nh} = Q\,\frac{\mu \mh}{\frac{4}{3} \pi \acore^3 \rho_c}.
\end{equation}
Numerically, this translates into
\begin{equation}\label{eq:xd_num}
X_d = 2.77\tdix{-12} \pfr{\acore}{0.1\micr}^{-3} \pfr{\rho_c}{2\gccc}^{-1} \pfr{Q}{0.01}.
\end{equation}
In our model setup, $X_d=1.69\tdix{-12}$.
%n_d [cm-3] = (1.4 n_H m_H) / (4/3 π \acore^3 rho_c) * Q
%= 2.79e-8 (n_H/1e4) (\acore/0.1mic)^-3 (rho_c/2gcm-3)^-1 (Q/1e-2)       
Given the above, the freeze-out timescale can be written as
\begin{equation}\label{eq:taufo_x0}
\taufo = x_0 \mamu^{1/2},
\end{equation}
where
\begin{equation}\label{eq:x0_def}
x_0^{-1} = \frac{3}{4}Q\frac{\mu \nh \mh}{\rho_c}\frac{\agr^2}{\acore^3}S\sqrt{\frac{8\kb T}{\pi\mh}}.
\end{equation}
Numerically, one obtains:
\begin{equation}\label{eq:x0}
%x_0 [{\rm yr\, amu^{-1/2}}] = {} 7.91\tdix{4}
\begin{aligned}
x_0 [{\rm yr}] = {} 7.91\tdix{4}
\pfr{\nh}{10^4\ccc}^{-1}
\pfr{Q}{0.01}^{-1}\\
\times\pfr{T}{10{\rm K}}^{-1/2}
\pfr{\agr}{0.1\micr}^{-2}
\pfr{\acore}{0.1\micr}^3
\pfr{\rhogr}{2\gccc}.&
\end{aligned}
\end{equation}
Incidentally, with our model setup parameters, the value of $x_0$ happens to be also {7.91\tdix{4} yr}. %/amu$^{0.5}$.

The freeze-out timescale for species such as N$_2$ or HCN is thus $\taufo\approx0.4$~Myr, and for atomic nitrogen, this is 0.3 Myr, in agreement with \rfig{depletion} showing the results of a constant-density model {(\nh=\dix{4}\ccc)}, but including adsorption. During the isothermal evolution of a starless core, the freeze-out timescale depends on the various quantities as:
\begin{equation}\label{eq:dtaufo}
\delta\taufo/\taufo = 1/2 (\delta \mamu/\mamu) - 2\delta \agr/\agr - \delta n_d/n_d.
\end{equation}
In the constant-density model of \rfig{depletion}, the adsorption timescale depends only on $m$ and \agr\ as
\begin{equation}\label{eq:dtaufo_3}
\delta\taufo/\taufo = 1/2 (\delta \mamu/\mamu) - 2\delta \agr/\agr.
\end{equation}
{We note that grain coagulation---which would decrease the grain surface area, and thus increase the freeze-out timescale (see Fig.2 of Flower et al 2005)---is not taken into account in this simple analysis.}
Hence, for a given species, the timescale evolves due to the increase in \agr, which is $\approx 10-15\%$. Now, for the \foun\ and \fifn\ variants of a given species, $\delta\mamu=1$, and at any instant, the relative difference in the freeze-out timescale reduces to
\begin{equation}\label{eq:dtaufo_2}
\delta\taufo/\taufo = 1/2 (\delta \mamu/\mamu) = 1/(2\mamu),
\end{equation}
or 1.8\% for HCN and \ce{N2}, and twice larger for N atoms.

At constant density, the number of a gas-phase species decreases due to adsorption as $n=n_0\exp(-t/\taufo)$ and the isotopic ratio of a \foun-species of mass $\mamu_{14}$ decreases with time as
\begin{equation}
\rrr(t) = \frac{n_{14}}{n_{15}} = \rrr_0 \exp[-t/(2 x_0 \mamu_{14} \sqrt{\mamu_{15}})].
\end{equation}
Therefore, differential depletion of \foun- and \fifn- variants will produce significant fractionation after a typical timescale (see \req{tauR})
\begin{equation}
\label{eq:taufr}
\taufr = 2 x_0 \mamu_{14} \sqrt{\mamu_{15}} \approx 2 x_0 \mamu^{3/2} = 2 \mamu\taufo,
\end{equation}
where we approximated $\mamu \approx \mamu_{14} \approx \mamu_{15}$. For our model parameters, the timescale for depletion-driven fractionation are \taufr=8.3~Myr for N atoms, and 23.6~Myr for \ce{N2} or HCN molecules.

%As can be seen in the right panel of \rfig{depletion}, the relative increase of \agr\ during gravitational collapse is modest ($\approx 10\%$), while $\delta n_d/n_d = \delta \nh/\nh \sim \delta t/\tauff$ during collapse (see \req{xd}). Thus, for a given species (fixed \mamu), the freeze-out timescale shortens as $\delta\taufo/\taufo = -\delta t/\tauff$.

Those numbers are a factor of 2.7 smaller in the constant-density models of \cite{loison2019}, who adopted $\rho_c=3$\gccc, \agr=0.1\micr, \nh=4\tdix{4}\ccc, and $Q=0.01$, for which {$x_0=2.96\tdix{4}$ yr}. With these parameters, the fractionation timescales become \taufr=3.1 and 8.8~Myr for N and HCN, respectively. One thus expects that, after a time 0.4 Myr in the models of \cite{loison2019} who considered an isotopic ratio of $\rrr_0=\rtot=440$, the relative decrease in \rrr\ should be $\delta \rrr/\rrr_0=0.4/5.1\approx5\%$ for HCN, or an isotopic ratio HCN/HC\fifn$\approx 390$. This is in good agreement for HCN (their Fig. 2), while for HNC their model leads to a ratio of $\approx 370$. We also note that, for atomic N, the relative variation is $\approx13\%$ while these authors report variations by several tens of percent which cannot be explained by differential depletion alone.

{From \req{x0}, we notice that $x_0$ depends on \agr\ through the ratio $\agr/\acore$ which is itself determined by the ratio of the core and mantles properties (compositions, mass density) to the power 1/3 (see \req{agr}). Hence this ratio is not expected to vary significantly (as already noticed from \rfig{depletion}). Also, the mass density of the refractory core is not expected to take values significantly different from our adopted value of 2\gccc. Thus one finds that $x_0\sim \acore/(Q\nh)$. Grain growth would thus increase the freeze-out and the fractionation timescales, as expected since the surface of grains per unit volume would decrease. However, higher values of the density and dust:gas mass ratio, such as toward the midplane of protoplanetary disks \citep{williams2014}, both shorten the fractionation timescale.}

\section{Dissociation of \ce{N2} and isotopologs by \ce{He+}}
\label{app:govers}

\begin{table}
	\centering
	\caption{The rates for the ionization and dissociation channels of the reaction \ce{N2 + He+}. The rate is computed as in \req{rate}.}
	\label{tab:govers}
	\begin{tabular}{ll lll}
		\toprule
		Reaction                               & $\alpha$ & $\beta$ & $\gamma$ &  \\
		                                       & \cccs    &  &K\\
		\midrule
		Default UGAN                           &  \\
		\ce{N2 +  He+ ->  N+  + N + He}        & 7.92e-10 & 0.0     & 0.0      &  \\
		\ce{^{15}NN + He+ -> ^{15}N+ + N + He} & 3.96e-10 & 0.0     & 0.0      &  \\
		\ce{^{15}NN + He+ -> N+ + ^{15}N + He} & 3.96e-10 & 0.0     & 0.0      &  \\
		\ce{N2 + He+  -> N2+  + He}            & 4.08e-10 & 0.0     & 0.0      &  \\
		\ce{^{15}NN +  He+ -> ^{15}NN+ + He}   & 4.08e-10 & 0.0     & 0.0      &  \\
		\midrule
		Modified rate                          &  \\
		\ce{N2 +  He+ ->  N+  + N + He}        & 7.20e-10 & 0.0     & 0.0      &  \\
		\ce{^{15}NN + He+ -> ^{15}N+ + N + He} & 2.45e-10 & 0.0     & 0.0      &  \\
		\ce{^{15}NN + He+ -> N+ + ^{15}N + He} & 2.45e-10 & 0.0     & 0.0      &  \\
		\ce{N2 + He+  -> N2+  + He}            & 4.80e-10 & 0.0     & 0.0      &  \\
		\ce{^{15}NN +  He+ -> ^{15}NN+ + He}   & 7.10e-10 & 0.0     & 0.0      &  \\
		\bottomrule
	\end{tabular}
\end{table}

\begin{figure}
	\centering
	\includegraphics[width=\hsize]{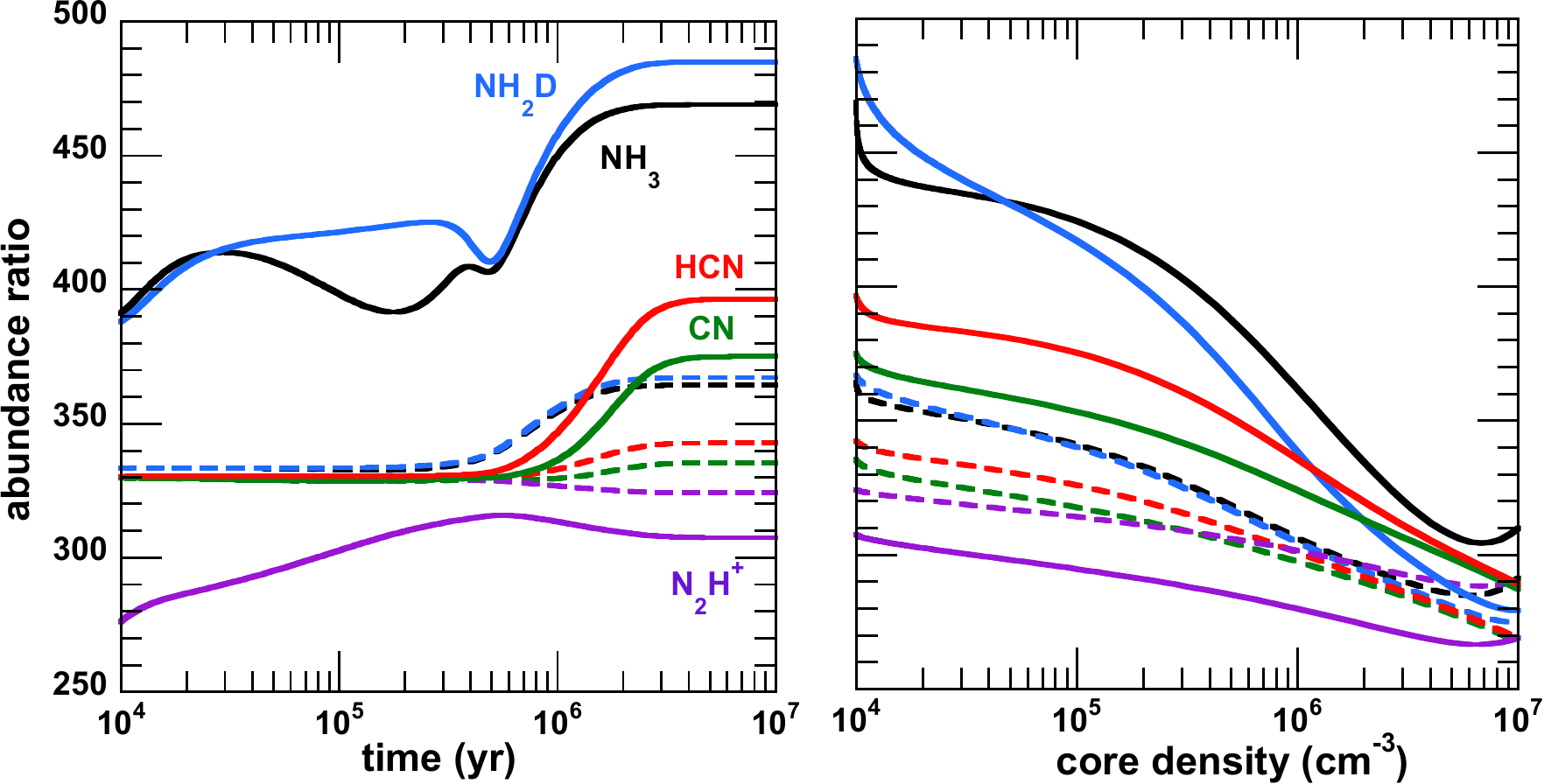}
	\caption{The effects on isotopic ratios of differing branching ratios, 1.5 and 0.69, respectively, for the reactions of $^{14}$N$^{14}$N and $^{14}$N$^{15}$N with \ce{He+} (\rsec{govers}). Results with different branching ratio (full curves) are compared to those obtained with both branching ratios equal to 1.5 (dashed curves). The \citetalias{roueff2015} set of fractionation reactions was used in these calculations.}
	\label{fig:govers}
\end{figure}

In \rfig{govers}, we illustrate, for the \citetalias{roueff2015} rates, the effects on isotopic ratios of differing branching ratios for the reactions of $^{14}$N$^{14}$N and $^{14}$N$^{15}$N with \ce{He+}, as discussed in \rsec{govers}. The modified rates for the two output channels of the \ce{N2+ + He+} reaction are listed in \rtab{govers}.

\subsection{Impact of cosmic-ray induced ultraviolet radiation}
\label{sec:app_photodiss}

\begin{figure}
	\centering
	\includegraphics[width=\hsize]{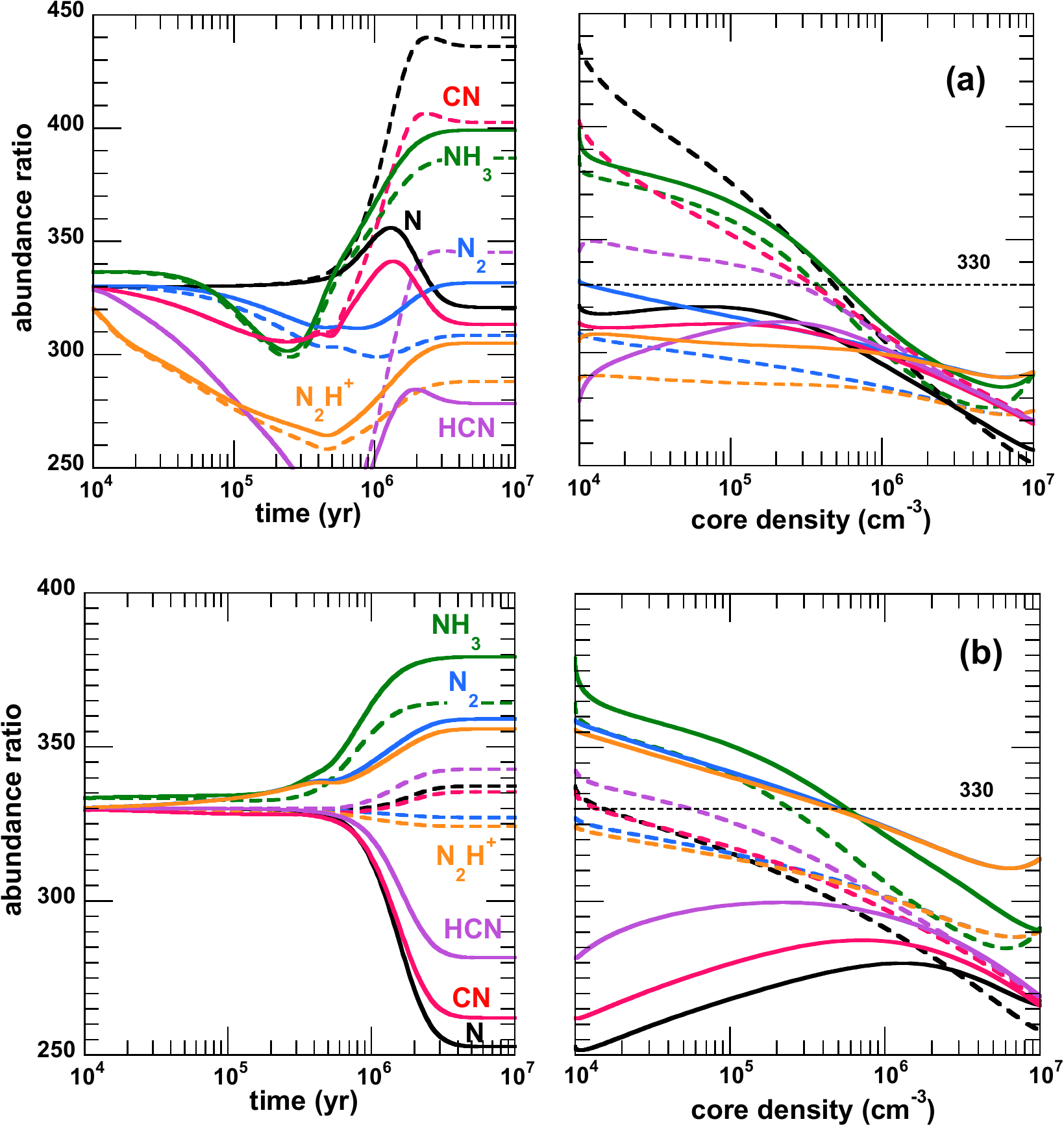}
	\caption{Influence of the higher rate of photodissociation of $^{15}$NN on the relative abundances of N- and $^{15}$N-containing species in the course of L-P collapse: using the fractionation reactions of  \citetalias{terzieva2000}, in panel (a), and  \citetalias{roueff2015}, in panel (b). Results with the enhanced rate are shown with full curves, while those using the default rate are in dashed.}
	\label{fig:CR}
\end{figure}

We consider the influence of differential dissociation of $^{14}$N$_2$ and \fifn\foun\ by the cosmic-ray induced ultraviolet radiation on the relative abundances of \foun- and \fifn-containing species, during their evolution to steady-state and in the course of gravitational collapse. As has been demonstrated \cite[][their Table 20]{heays2017}, the rates of photodissociation of N$_2$ and $^{15}$NN, by the cosmic-ray induced radiation, could differ by approximately a factor of three when self-shielding of the principal isotopic form, N$_2$, is significant. Following these authors, we adopt a rate of photodissociation of N$_2$ of $39 \zeta_\hh$, while the unshielded rate is 120$\zeta_\hh$ (see their footnote \textit{c}). In Fig.~\ref{fig:CR}, we thus compare results obtained with either the same rate or a three-times higher rate for $^{15}$NN. When the latter value is used, molecular and atomic nitrogen are respectively depleted and enriched in $^{15}$N, with consequences for the other nitrogen-containing species that depend on their routes of formation.

%We here study the impact of the differential dissociation of $^{14}$N$_2$ and \fifn\foun\ by the cosmic-ray-induced ultraviolet radiation, on the relative abundances of \foun- and \fifn-containing species during the collapse.

%As was demonstrated by \cite[][their Table 20]{heays2017}, the rates of photodissociation of N$_2$ and $^{15}$NN, by the cosmic-ray-induced radiation, differ by approximately a factor 3, owing to the greater degree of self-shielding of the principal isotopic form, N$_2$. Fig.~\ref{fig:CR} illustrates the effect of the higher rate of photodissociation of $^{15}$NN on the relative abundances during collapse of selected N- and $^{15}$N-containing species. As might have been anticipated, molecular nitrogen is depleted, and atomic nitrogen enriched, in $^{15}$N, with consequences for the other nitrogen-containing species that depend on their routes of formation.

\end{document}